\documentclass[fleqn,usenatbib]{mnras}

\usepackage[T1]{fontenc}
\usepackage{ae,aecompl}

\usepackage{graphicx}	% Including figure files
\usepackage{amsmath}	% Advanced maths commands
\usepackage{amssymb}	% Extra maths symbols
\usepackage{longtable}	% Extra maths symbols
\newcommand{\BF}[1]{}

\title[AGN and outflows in jellyfish galaxies]{GASP XIX: AGN and their outflows at the center of jellyfish galaxies}

\author[M. Radovich et al.]{
	Mario Radovich$^{1}$\thanks{E-mail: mario.radovich@inaf.it},
	Bianca Poggianti$^{1}$,
	Yara L. Jaff\'e$^{2}$,
	Alessia Moretti$^{1}$,
	\newauthor
	Daniela Bettoni$^{1}$,
	Marco Gullieuszik$^{1}$,
	Benedetta Vulcani$^{1}$,
	and Jacopo Fritz$^{3}$
	\\
	$^{1}$INAF- Osservatorio astronomico di Padova, Vicolo Osservatorio 5, IT-35122 Padova, Italy\\
	$^{2}$Instituto de F\'isica y Astronom\'ia, Facultad de Ciencias, Universidad de Valpara\'iso, Avda. Gran Breta\~na 1111, Casilla 5030, Valpara\'iso, Chile \\
	$^{3}$Instituto de Radioastronom\'ia y Astrof\'isica, UNAM, Campus Morelia, A.P. 3-72, C.P. 58089, Mexico}

\date{Accepted 2019 March 15. Received 2019 March 15; in original form 2019 February 5}

\pubyear{2019}

\begin{document}
\label{firstpage}
\pagerange{\pageref{firstpage}--\pageref{lastpage}}
\maketitle

\begin{abstract}

The GASP survey, based on MUSE data, is unveiling the properties of the gas in the so-called "jellyfish" galaxies: these are cluster galaxies with spectacular evidence of gas stripping by ram pressure. In a previous paper, we selected the seven GASP galaxies with the most extended 
tentacles of ionized gas, and based on individual diagnostic diagrams concluded that at least five of them present clear evidence for an Active Galactic Nucleus.
Here we present a more detailed analysis of the emission lines properties in these galaxies. Our comparison of several emission line ratios with both AGN and shock models show that photoionization by the AGN is the dominant ionization mechanism. This conclusion is strengthened by the analysis of $\rm H\beta$ luminosities, the presence of nuclear iron coronal lines and extended ($>10$ kpc) emission line regions ionized by the AGN in some of these galaxies. From emission line profiles, we find the presence of outflows in four galaxies, and derive mass outflow rates, timescales and kinetic energy of the outflows.  
\end{abstract}


%%%%%%%%%%%%%%%%%%%%%%%%%%%%%%%%%%%%%%%%%%%%%%%%%%

%%%%%%%%%%%%%%%% BODY OF PAPER %%%%%%%%%%%%%%%%%%

\begin{table*}
	\begin{tabular}{cccccccccccccc}
		\hline\hline
		id & cluster & z$_{\rm cl}$ & scale  & RA & DEC &  class  & seeing & $r_{\rm NLR}$  &  $r_{\rm AGN}$ & $\log L_{\rm [OIII]}$& $\log L_{\rm [OIII]}^{\rm corr}$ \\
		& &   & kpc/$\arcsec$ & &  & & $\arcsec$&kpc & kpc & erg s$^{-1}$ & erg s$^{-1}$\\
		\hline
		JO135 & A3530 & 0.05480   & 1.07& 12 57 04.2  &  -30 22 30.0  & AGN& 0.73 & 0.8 & 3.7 & 41.24 $\pm$ 0.03  & 42.16 $\pm$ 0.03 \\
        JO175 & A3716 & 0.04599 & 0.90& 20 51 17.6 & -52 49 21.8    & SF& 1.05 & & \\
        JO194 & A4059 & 0.04877  &  0.95& 23 57 00.7 & -34 40 50.4   & LINER& 0.82&  0.6 & 0.7 &  38.25 $\pm$ 0.01  & 39.30 $\pm$ 0.01  \\ 
        JO201 & A85 & 0.05568   & 1.08& 00 41 30.3  &  -09 15 45.9   & AGN  & 0.99& 1.1 & 3.7 &  41.90 $\pm$ 0.02  & 42.20 $\pm$ 0.02  \\
        JO204 & A957 & 0.04496 &  0.88 & 10 13 46.8  &  -00 54 50.9  & AGN & 0.87 & 0.7 & 1.8 & 40.29 $\pm$ 0.02  & 41.25 $\pm$ 0.02 \\
        JO206 & IIZW108 & 0.04889 &  0.96& 21 13 47.4  &  +02 28 34.1  & AGN& 1.14 & 1.0 & 2.6 &  40.78 $\pm$ 0.02  & 41.41 $\pm$ 0.02\\
        JW100 & A2626& 0.05509 &  1.07 & 23 36 25.0  &   +21 09 02.5   & AGN & 1.09 &  1.3 & 2.8 & 40.37 $\pm$ 0.02  & 40.79 $\pm$ 0.02\\
		
		\hline    
	\end{tabular} 
	\caption{The table shows: the galaxy ID, the host cluster name, redshift and scale,  the coordinates of the central spaxel,  the seeing, the classification (AGN/LINER/SF) assigned in \citet{2017Natur.548..304P} and,  for galaxies classified as AGN or LINER, the estimated AGN sizes and the observed and dereddened [OIII] $\lambda$5007 luminosities within $r_{\rm NLR}$.\label{tab:props}}
\end{table*}

\section{Introduction}

It is now widely accepted that there is a strong connection between the presence of an Active Galactic Nucleus (AGN) and the host galaxy properties, based both on cosmological models and observational results from wide-field surveys 
\citep[see e.g.][and refs for a review]{2014ARA&A..52..589H}. However, the way this interaction occurs is still unclear, and it may actually be the outcome of a wide range of different physical processes (e.g. merging, bars).  A major improvement in our understanding of the complex environment around AGN is given by the availability of Integral Field Spectroscopy (IFU), allowing to map emission line fluxes and kinematics tracing the AGN and its surroundings \citep[see e.g.][]{2018A&A...619A..74V, 2019MNRAS.484..252I, 2019A&A...622A.146M}.

An important issue is the effect of the environment 
on the presence of the AGN: it is still debated  \citep[see e.g.][and refs]{2017A&A...599A..83M} whether or not  a dense galaxy environment such as in galaxy clusters has any effect on the presence of AGN. Early spectroscopic studies  \citep{1985ApJ...288..481D} suggested that the fraction of AGN in clusters ($\sim$1\%) is significantly lower than in a field environment ($\sim$5\%). Later studies based on X-ray data \citep{2013MNRAS.429.1827P} did not however confirm this result, showing comparable fractions of AGN in cluster and field, though there may be an effect related to distance from the cluster centre \citep{2014MNRAS.437.1942E}. 
In this context, \citet{2018MNRAS.474.3615M} used semi--analytic galaxy evolution models to show that   both star formation and AGN can be triggered by the ram pressure as they move through the intracluster medium, in galaxies located at  distances from the cluster centre larger than the virial radius; at smaller distances, where the ram pressure is higher, models suggest that the gas is stripped from the galaxy and can't feed the AGN. 
\citet{2018MNRAS.476.3781R}  analyzed the role of the galactic magnetic field in the gas stripping using 3D magnetohydrodynamic simulations: they found that the magnetic field can contribute to generate a gas inflow to the central parts of the galaxies, triggering star formation and maybe feeding the AGN.

Conversely, the presence of the AGN may  impact the surrounding environment in many ways \citep[see e.g.][for a review]{2012ARA&A..50..455F}: in particular, they are able to drive outflows of ionized gas and impact on the galaxy environment on scales that may range from few kpcs \citep[see e.g.][]{2019MNRAS.482..194B} for the less luminous AGN to tens of kpc for the brightest AGN \citep{2014MNRAS.441.3306H}.

The {\em GAs Stripping Phenomena} (GASP) survey
 \citep[][P17a hereafter]{2017ApJ...844...48P} is aimed at studying with the MUSE Integral Field spectrograph on VLT the properties of the so-called {\em jellyfish} galaxies in clusters, whose {\em tentacles} of UV and optically bright material that make them similar to a jellyfish \citep{2010MNRAS.408.1417S} are thought to originate via ram-pressure stripping by the intra-cluster medium \citep{2014ApJ...781L..40E,2014MNRAS.445.4335F,2014MNRAS.442..196R,2016MNRAS.455.2028F}. \citet{2017Natur.548..304P} (P17b hereafter)  showed that at least five and possibly six of  seven galaxies with the strongest evidence of gas stripping and the most favourable conditions for ram pressure  \citep{2018MNRAS.476.4753J} host an AGN, suggesting a connection between ram pressure stripping and AGN triggering. 
In  P17b the [NII]$\lambda$6583/H$\alpha$ vs. [OIII]$\lambda$5007/H$\beta$ line ratios were used to select the most likely mechanism that ionized the gas: radiation from hot young stars in star-forming regions, from an AGN, a combination of them ({\em composite}), and either low-luminosity AGN or shocks (LINERs), using as reference the classification by \citet{2006MNRAS.372..961K}. As already shown in \citet{2019MNRAS.482.4466P}, adding other line ratio diagnostic diagrams such as [OI]$\lambda$6300/H$\alpha$ and [SII]$\lambda\lambda$6716,6731/H$\alpha$ vs. [OIII]$\lambda$5007/H$\beta$ \citep{1987ApJS...63..295V} can provide a more detailed description of the physical processes at work.
In this paper we expand the work by P17b and critically scrutinize those results using all three main diagnostic diagrams simultaneously and comparing observed line ratios with photoionization and shock models. Moreover, we inspect additional features such as coronal Fe lines and analyze separately the extended extranuclear AGN-powered emission regions. Finally, we discuss the presence and properties of outflows.

The paper is structured as follows. A short summary of the data and how they were analyzed is given in Sect.\ref{sec:Data}. 
In Sect.\ref{sect:analysis}, observed emission line ratios are compared with both photoionization and shock models, to  confirm that photoionization from the AGN is required to reproduce the line ratios and derive the best-fit model parameters. As further probes of the AGN, we estimate the maximum contribute to the observed H$\beta$ luminosity from shock models; in some cases, we detect the presence of high-ionization iron coronal lines (JO201 and JO135) and of  extended ($>$ 10 kpc) AGN-like emission lines (JO204 and JO135): this is used in  Sect.\ref{sect:ExtNucleus} to derive the number of ionizing photons that should be emitted by the AGN.
In  Sect.\ref{sect:Outflows} we analyze the [OIII]$\lambda$5007 line as a tracer of outflows around the AGN and derive their size, outflow rates, timescales and kinetic energy. Conclusions are given in Sect.~\ref{sect:conclusions}.

The cosmology concordance model was adopted: $H_0 = 70$ km s$^{-1}$ Mpc$^{-1}$, $\Omega_m = 0.3$, $\Omega_\Lambda = 0.7.$

\begin{figure*}
	\begin{tabular}{cc}
		JO135 & JO201\\
			\includegraphics[trim=20 0 50 40,clip,width=.45\linewidth]{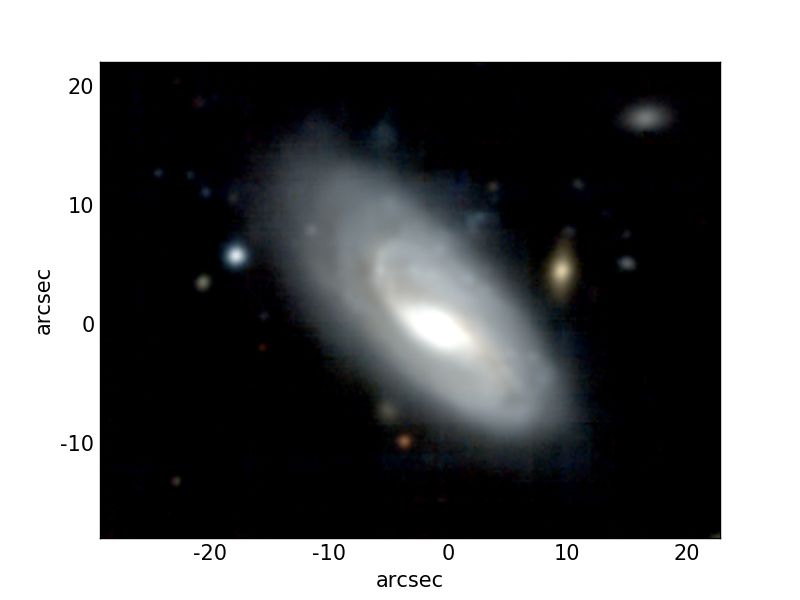} &
		\includegraphics[trim=10 0 10 40,clip,width=.5\linewidth]{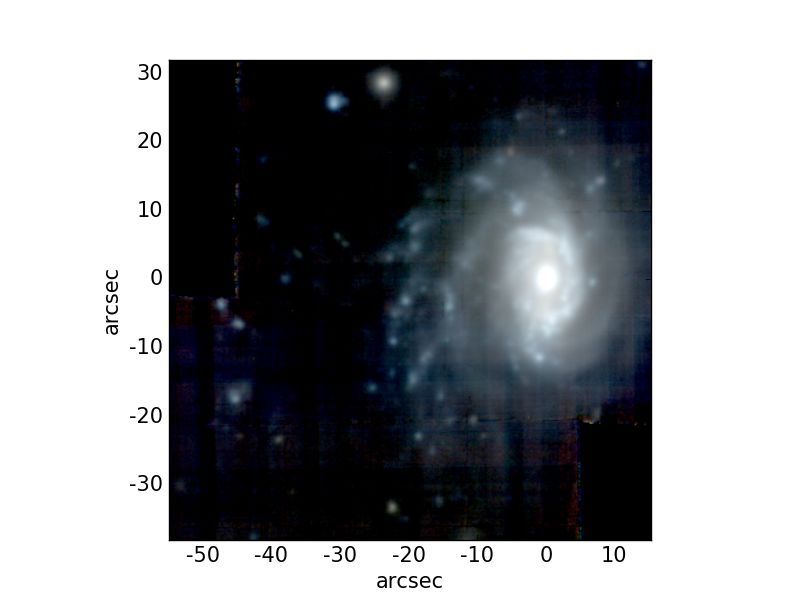}\\
		JO204 & JO206\\	
		\includegraphics[trim=50 0 50 40,clip,width=.45\linewidth]{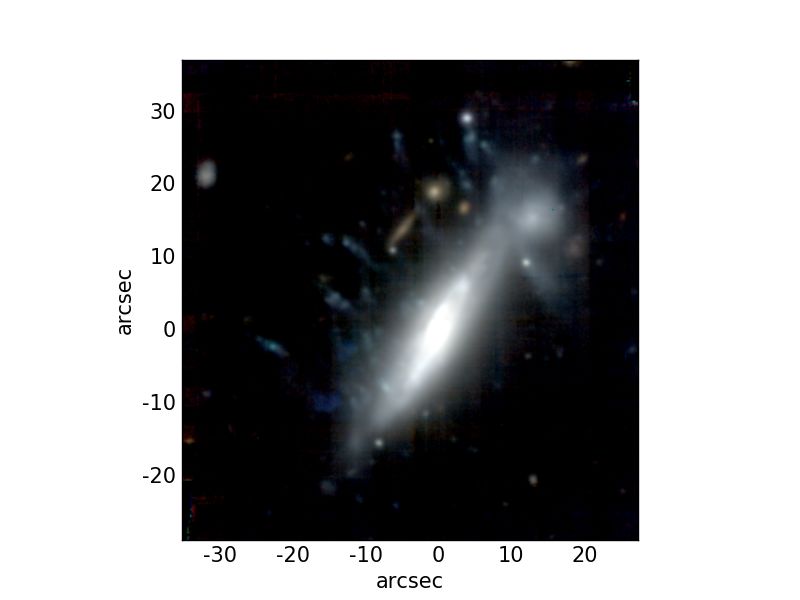}	&
		\includegraphics[trim=50 0 50 40,clip,width=.45\linewidth]{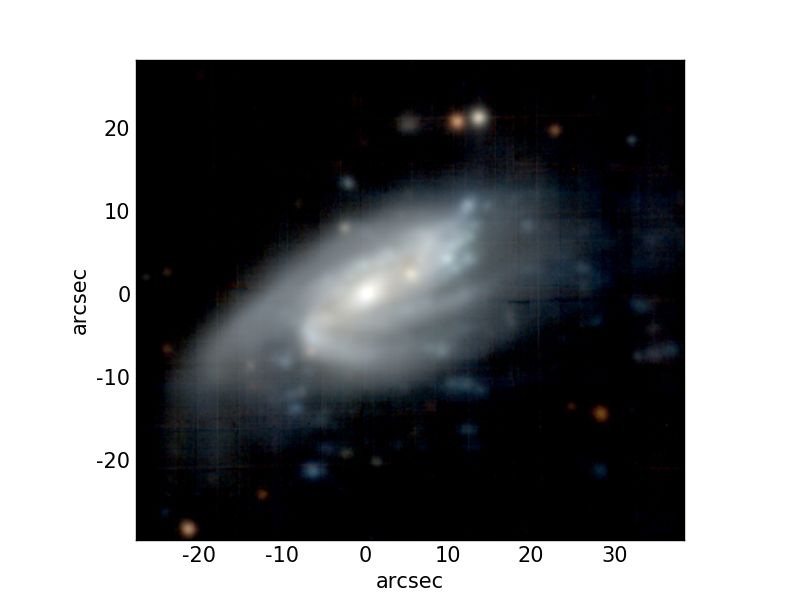}\\	
	\end{tabular} 
		\begin{tabular}{ccc}
			JW100 & JO194 & JO175\\	
		\includegraphics[trim=110 0 150 40,clip,width=.3\linewidth]{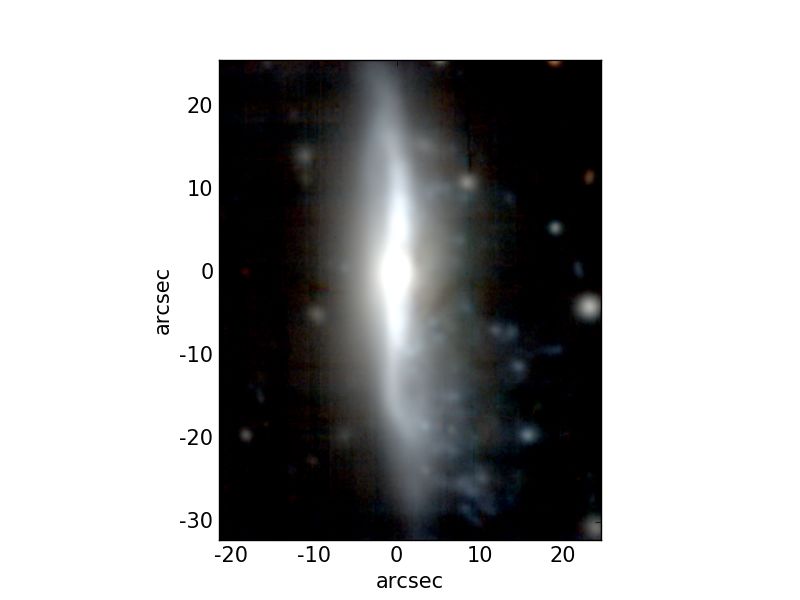}	&
		\includegraphics[trim=130 0 150 40,clip,width=.28\linewidth]{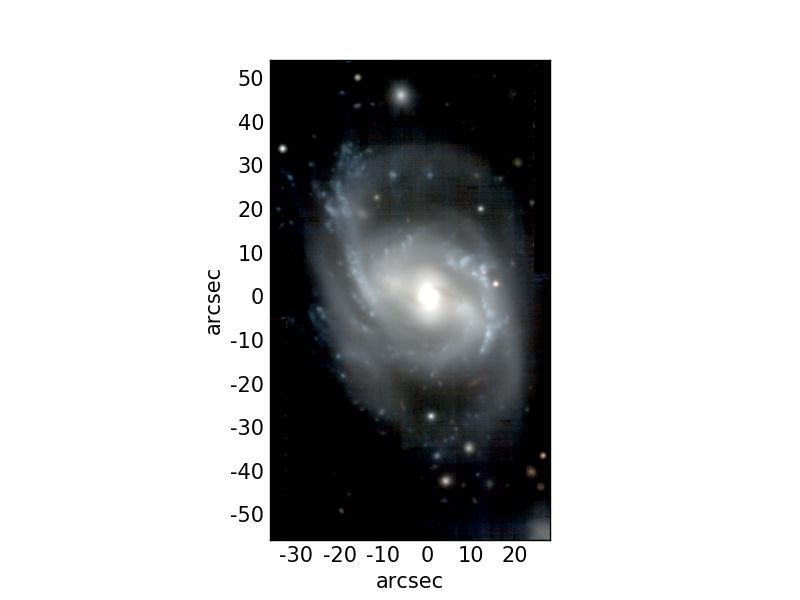}& 	
		\includegraphics[trim=110 0 130 40,clip,
		width=.32\linewidth]{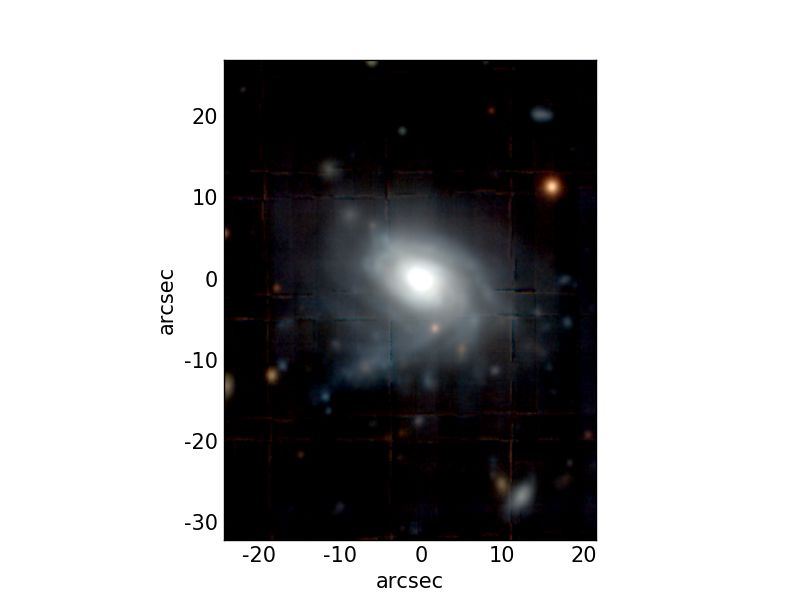}	\\	

	\end{tabular} 
	\caption{VRI images built from the MUSE cubes for the seven galaxies analyzed in the paper.\label{fig:VRI}}
\end{figure*}		

\begin{figure*}
	\begin{tabular}{cc}
	\includegraphics[width=.5\linewidth]{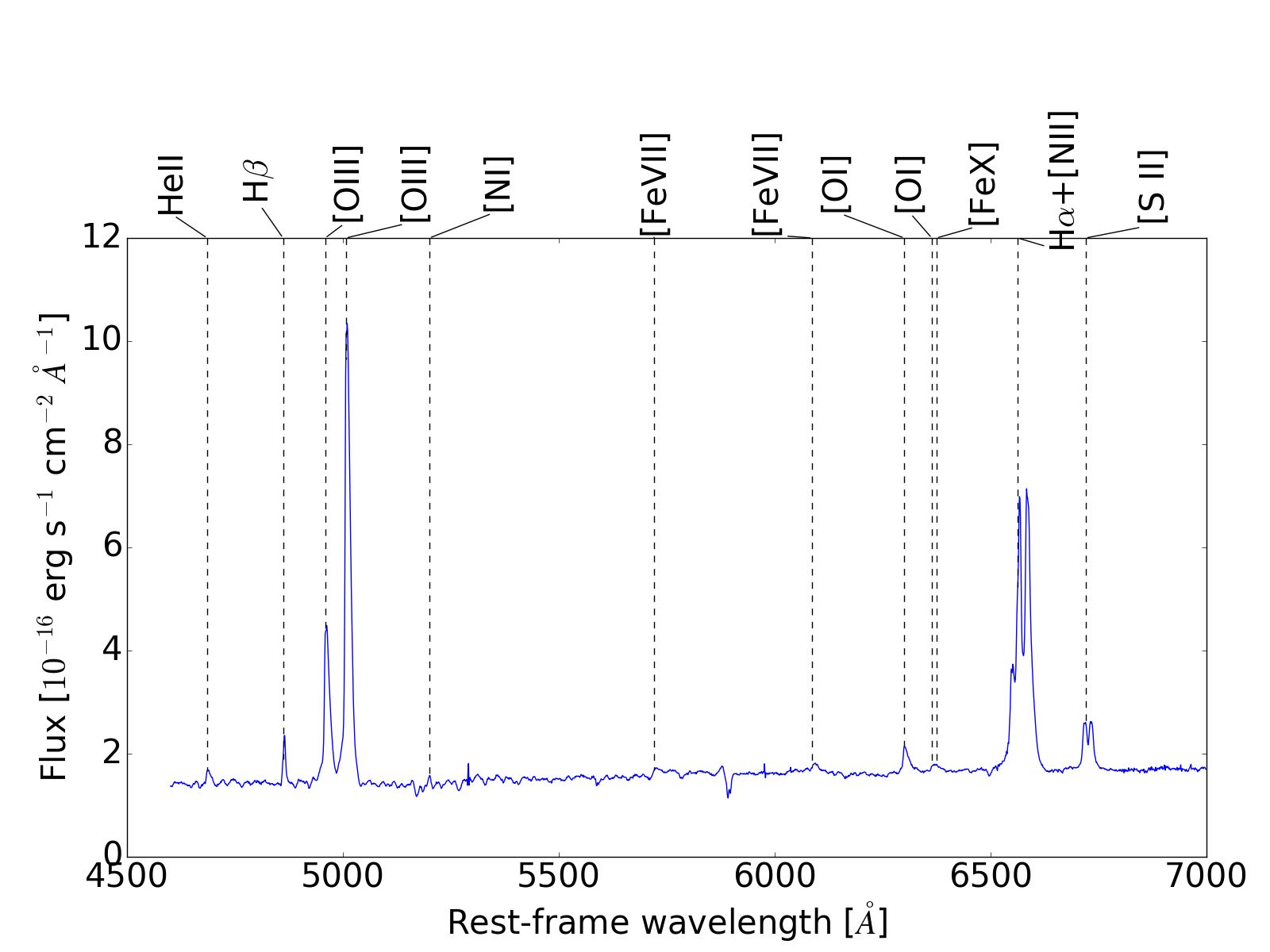}	&
	\includegraphics[width=.5\linewidth]{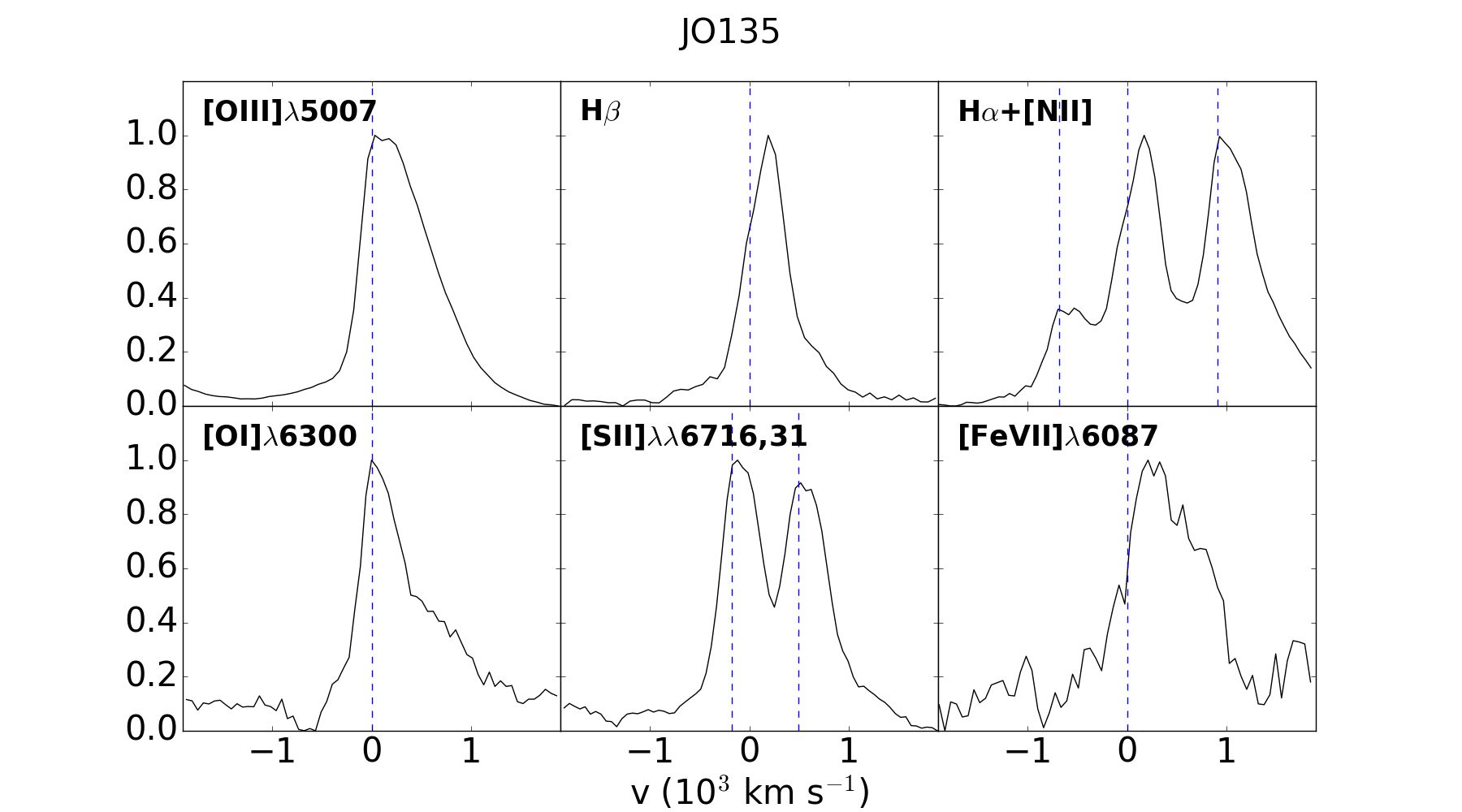} \\

	\includegraphics[width=.5\linewidth]{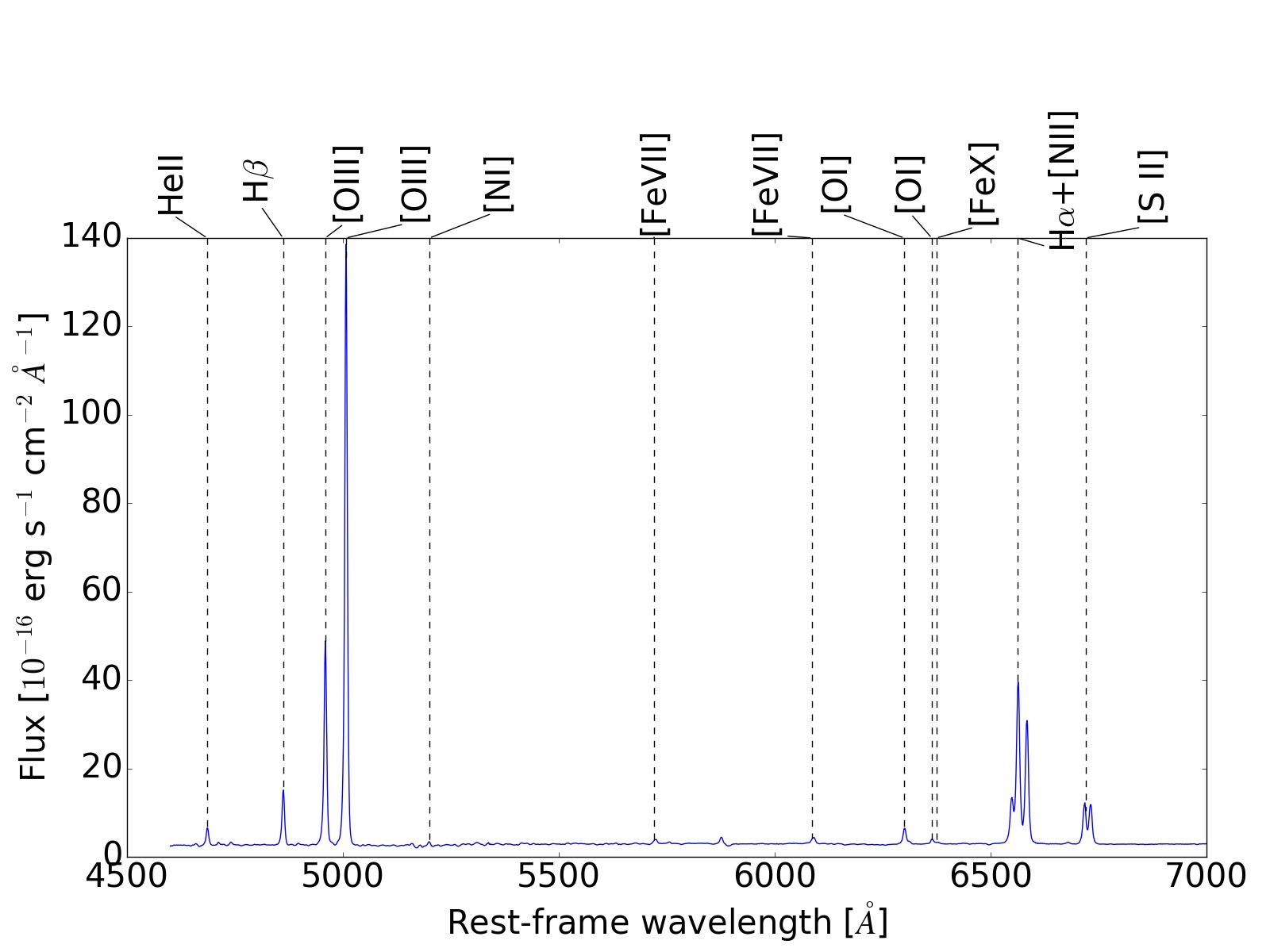}	&
	\includegraphics[width=.5\linewidth]{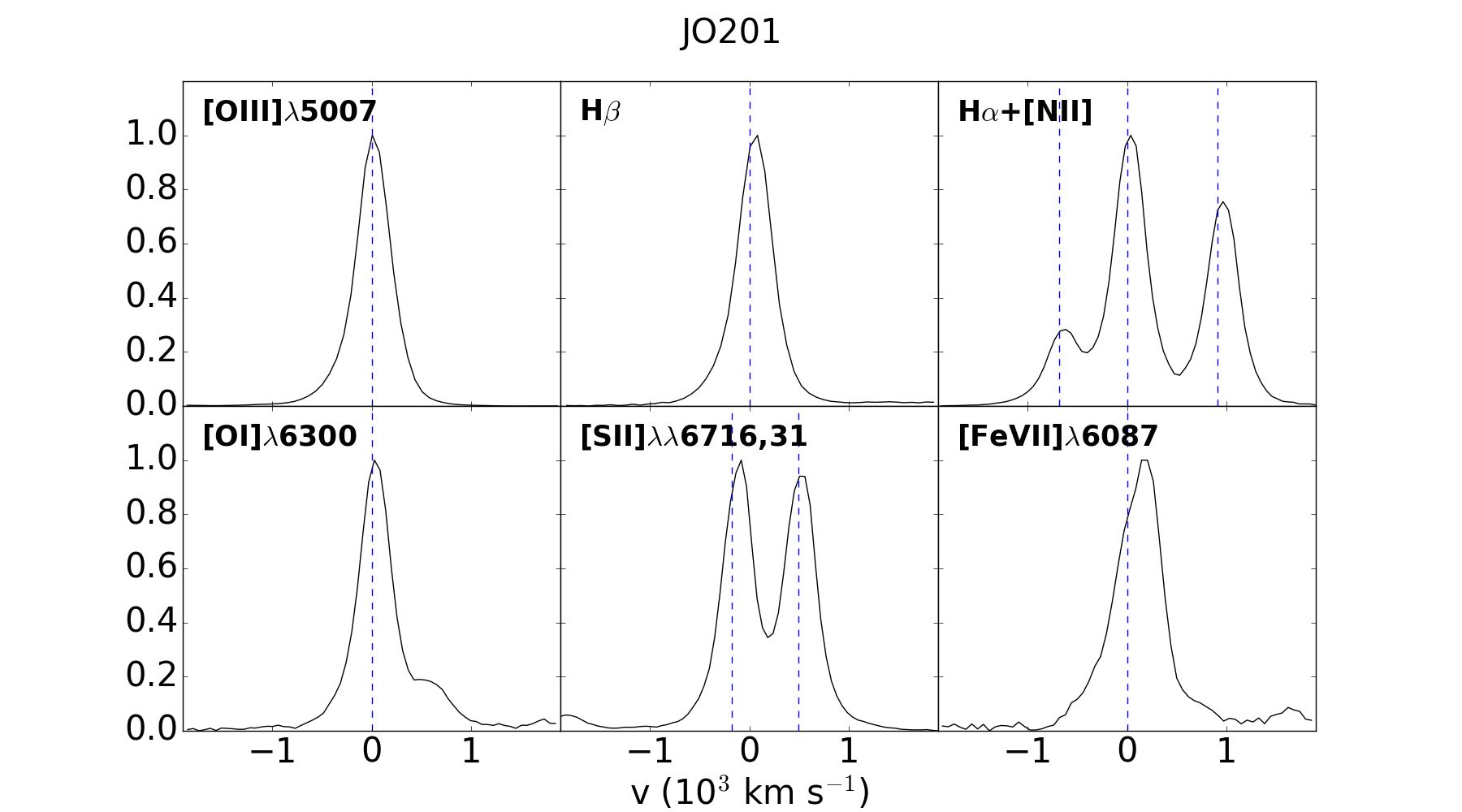} \\
	\includegraphics[width=.5\linewidth]{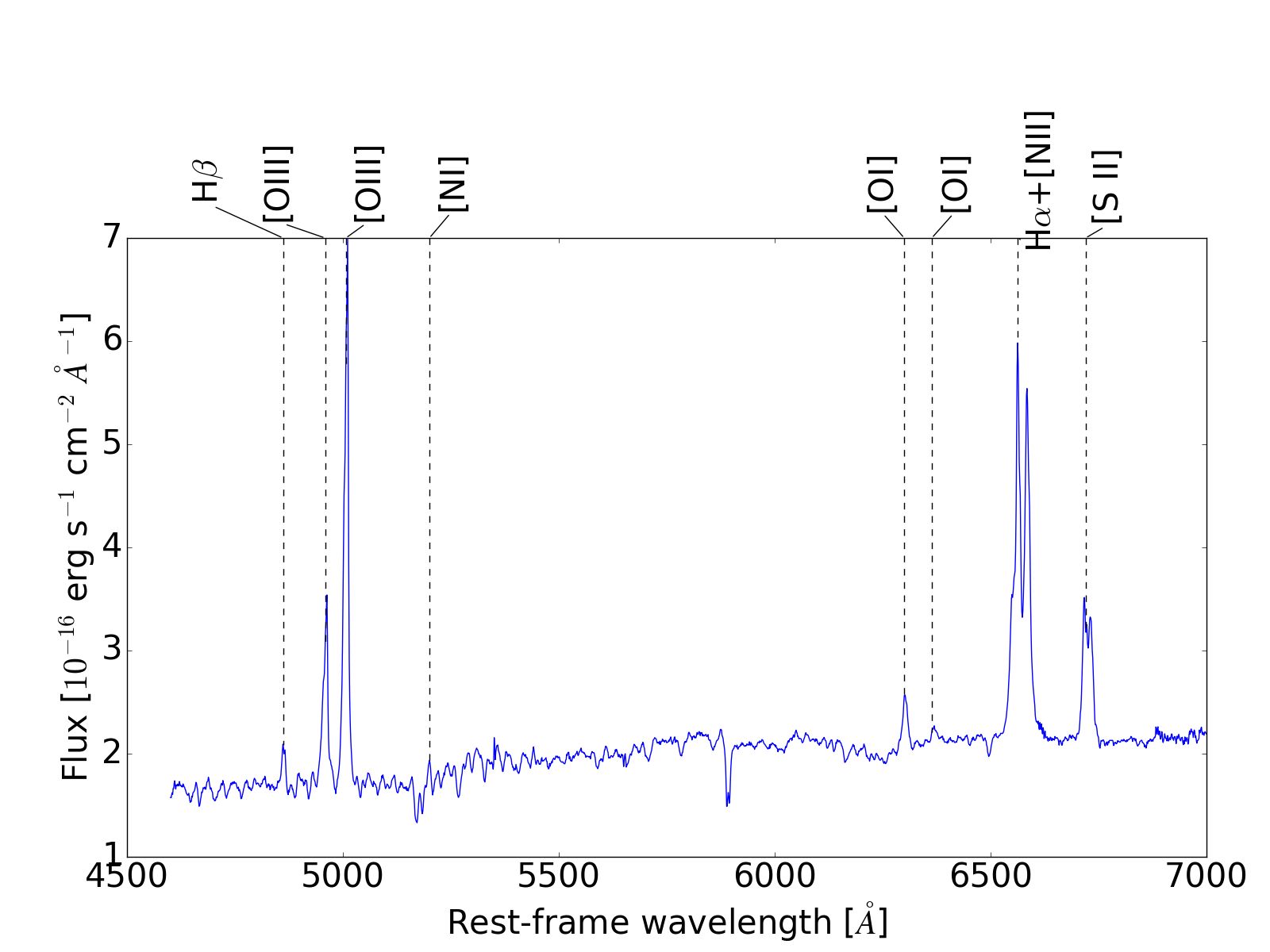}	&
	\includegraphics[width=.5\linewidth]{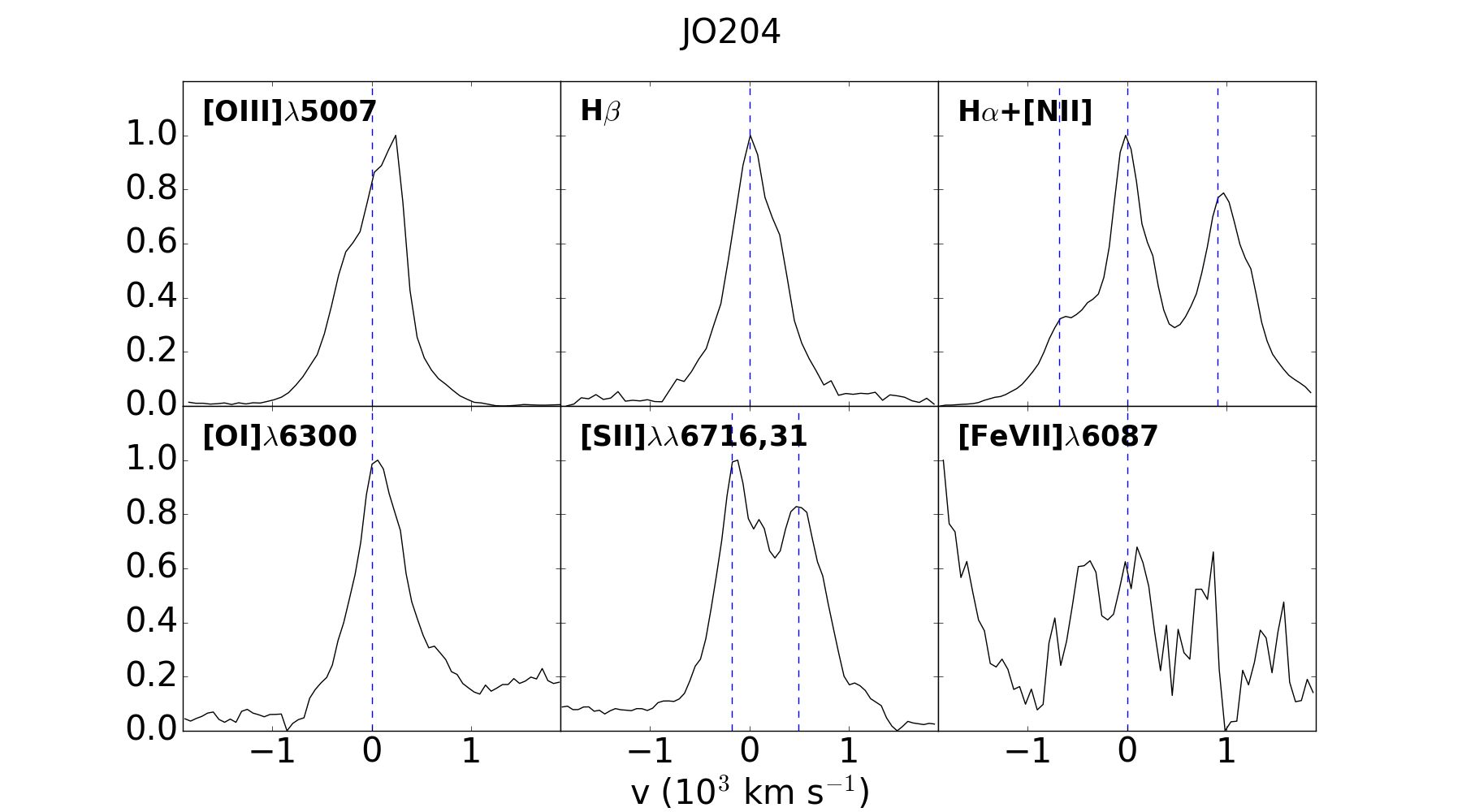} \\
	\end{tabular} 
	\caption{The plots show for the central spaxel of each galaxy: {\em left -} the uncorrected spectra with the identification of the main emission lines; {\em right -} the normalized  line profiles after the subtraction of the stellar underlying spectrum as described in the text: the dashed lines indicate the assumed galaxy velocity. A meaningful detection (SN > 3) of [FeVII]$\lambda$6087 is seen only in JO135 and JO201. \label{fig:spectra}}
\end{figure*}

\begin{figure*}
	\begin{tabular}{cc}		
	\includegraphics[width=.5\linewidth]{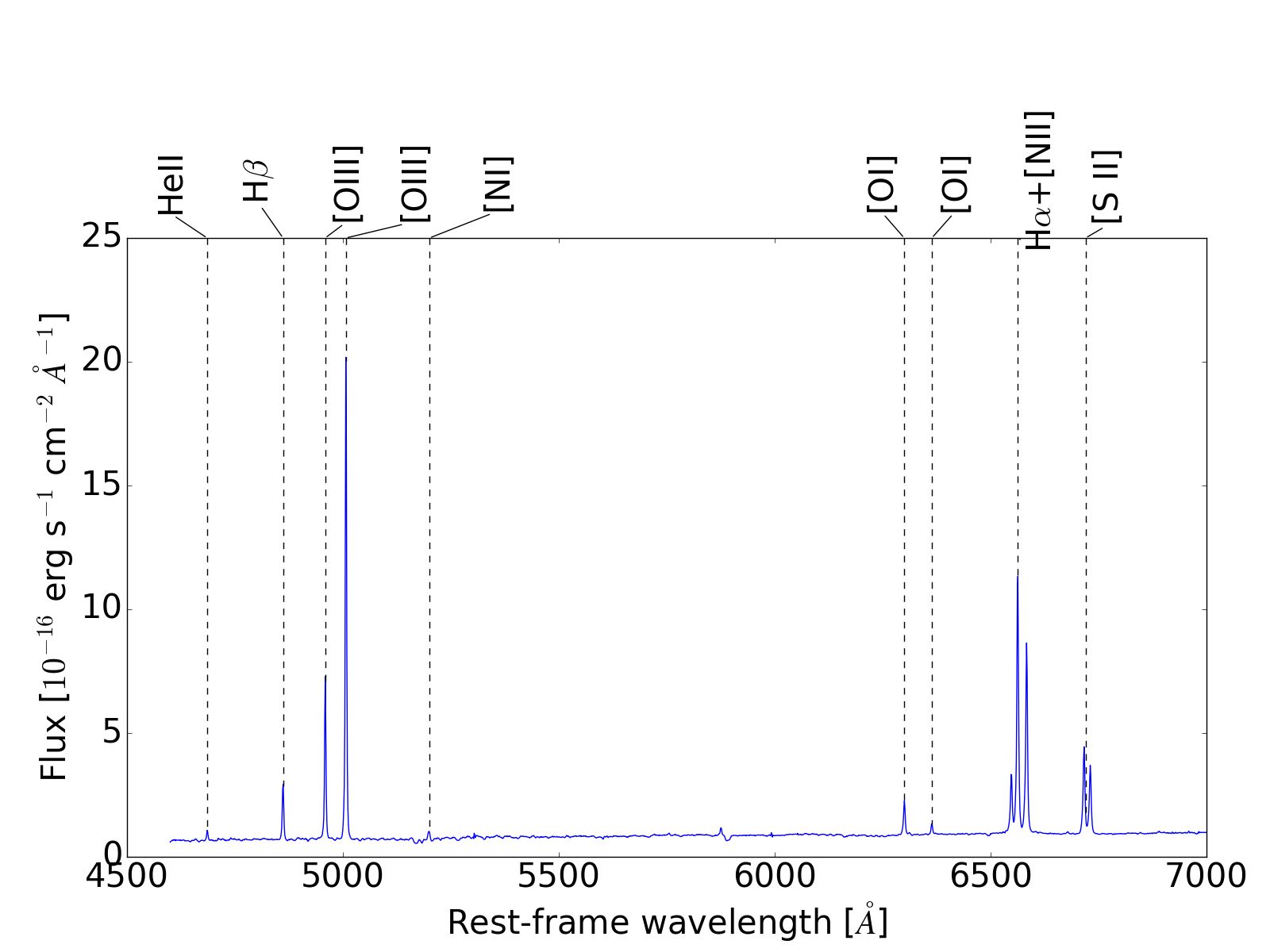}	&
	\includegraphics[width=.5\linewidth]{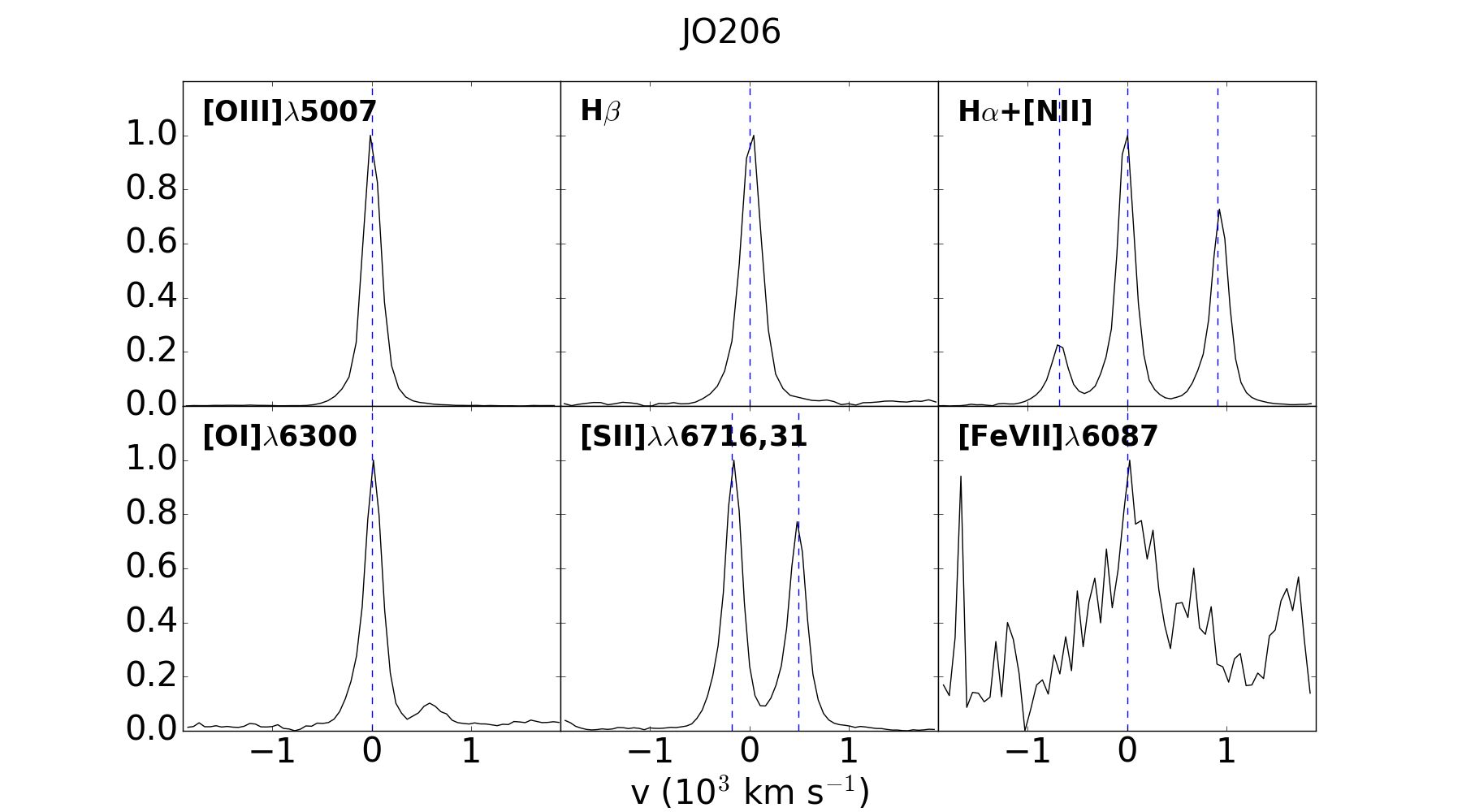} \\
	\includegraphics[width=.5\linewidth]{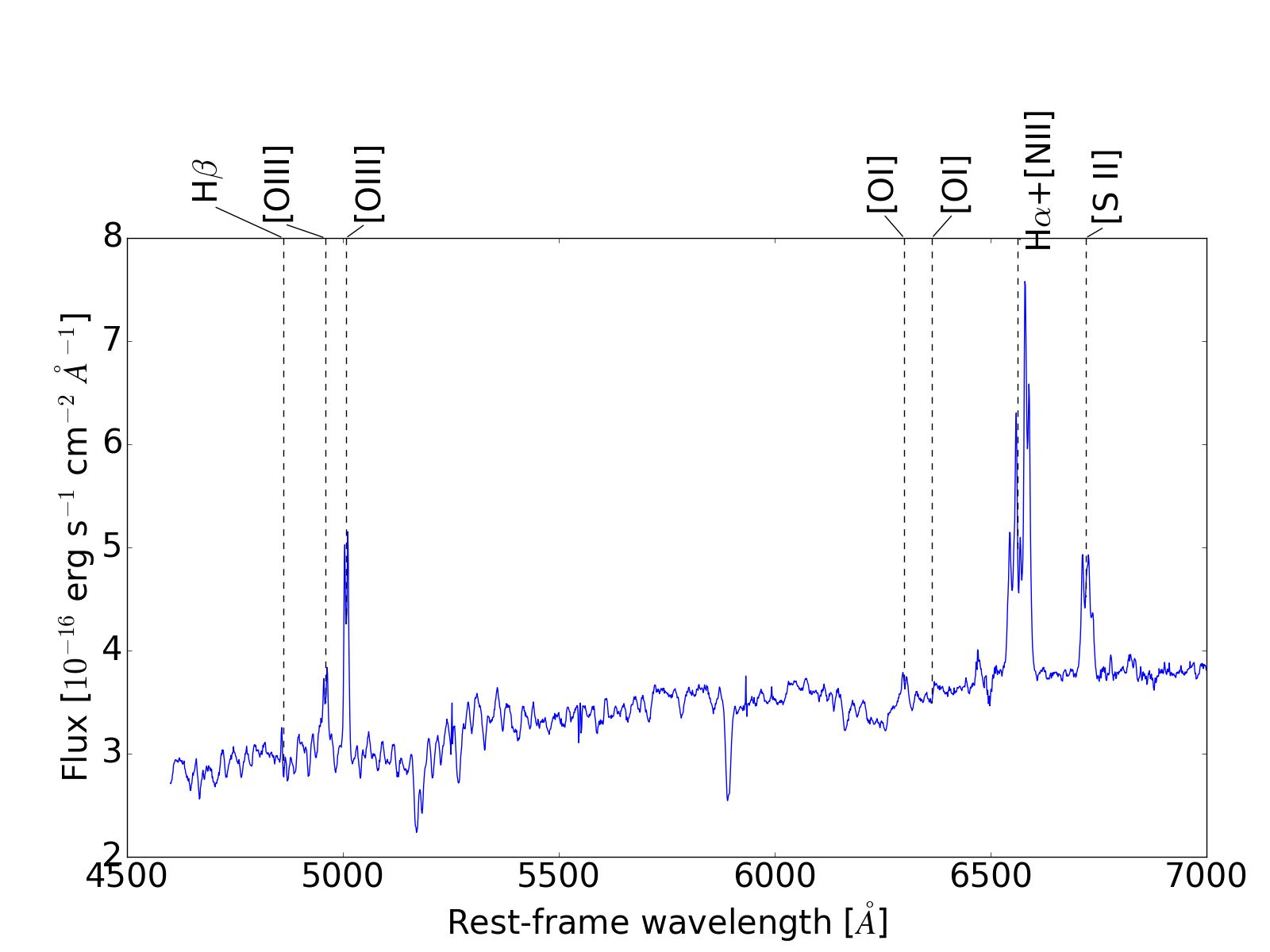}	&
	\includegraphics[width=.5\linewidth]{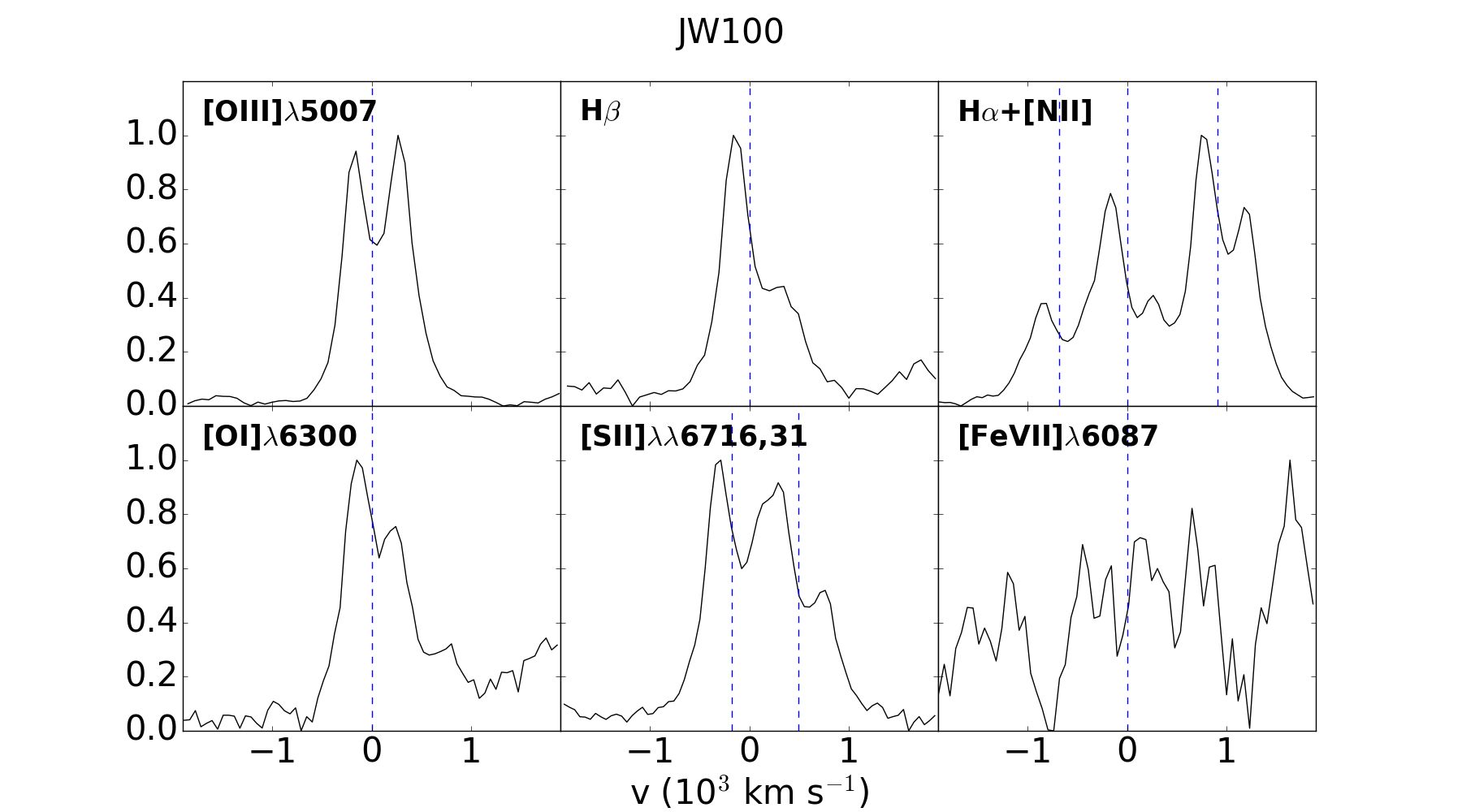} \\
	\includegraphics[width=.5\linewidth]{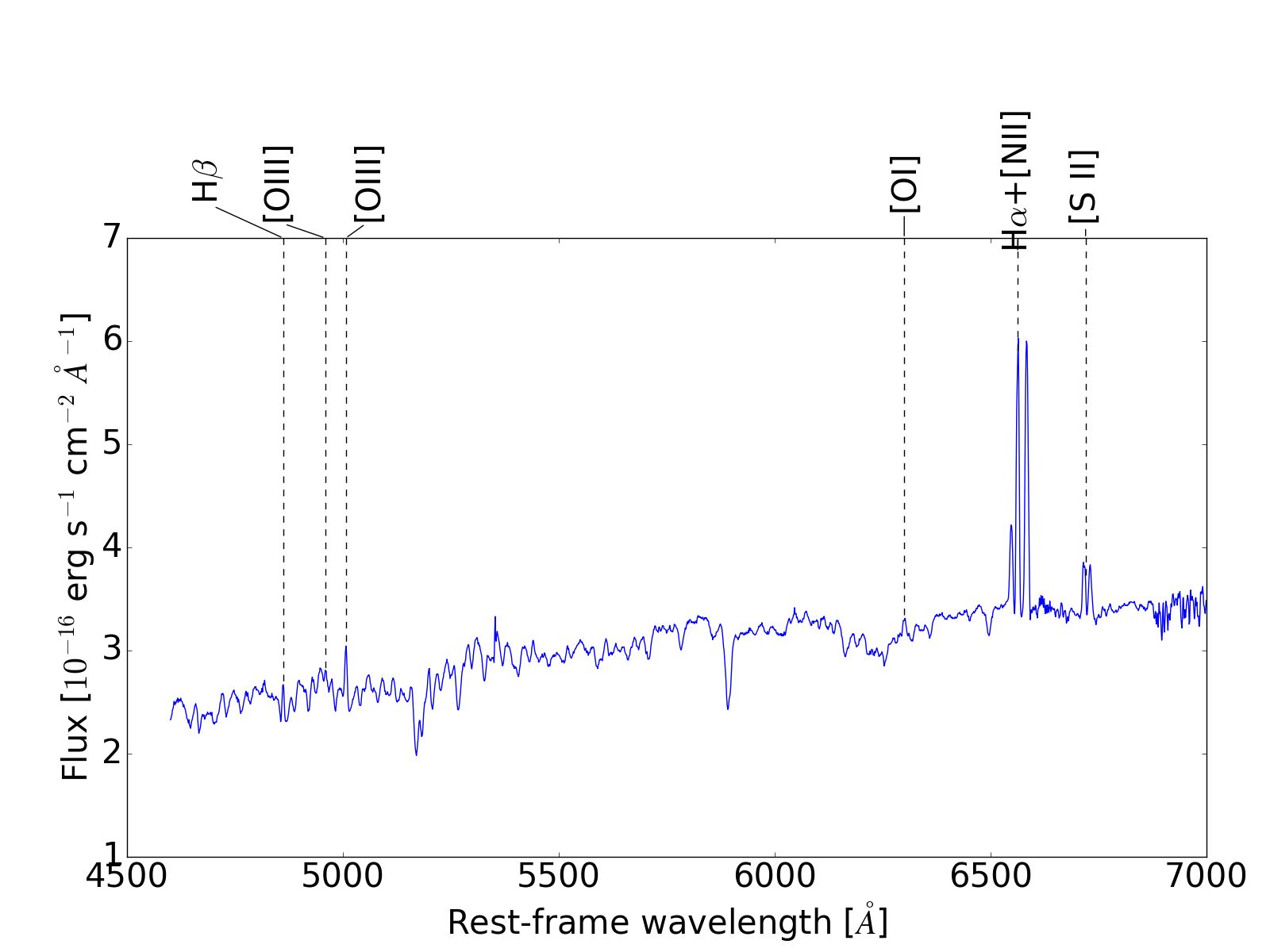}	&
	\includegraphics[width=.5\linewidth]{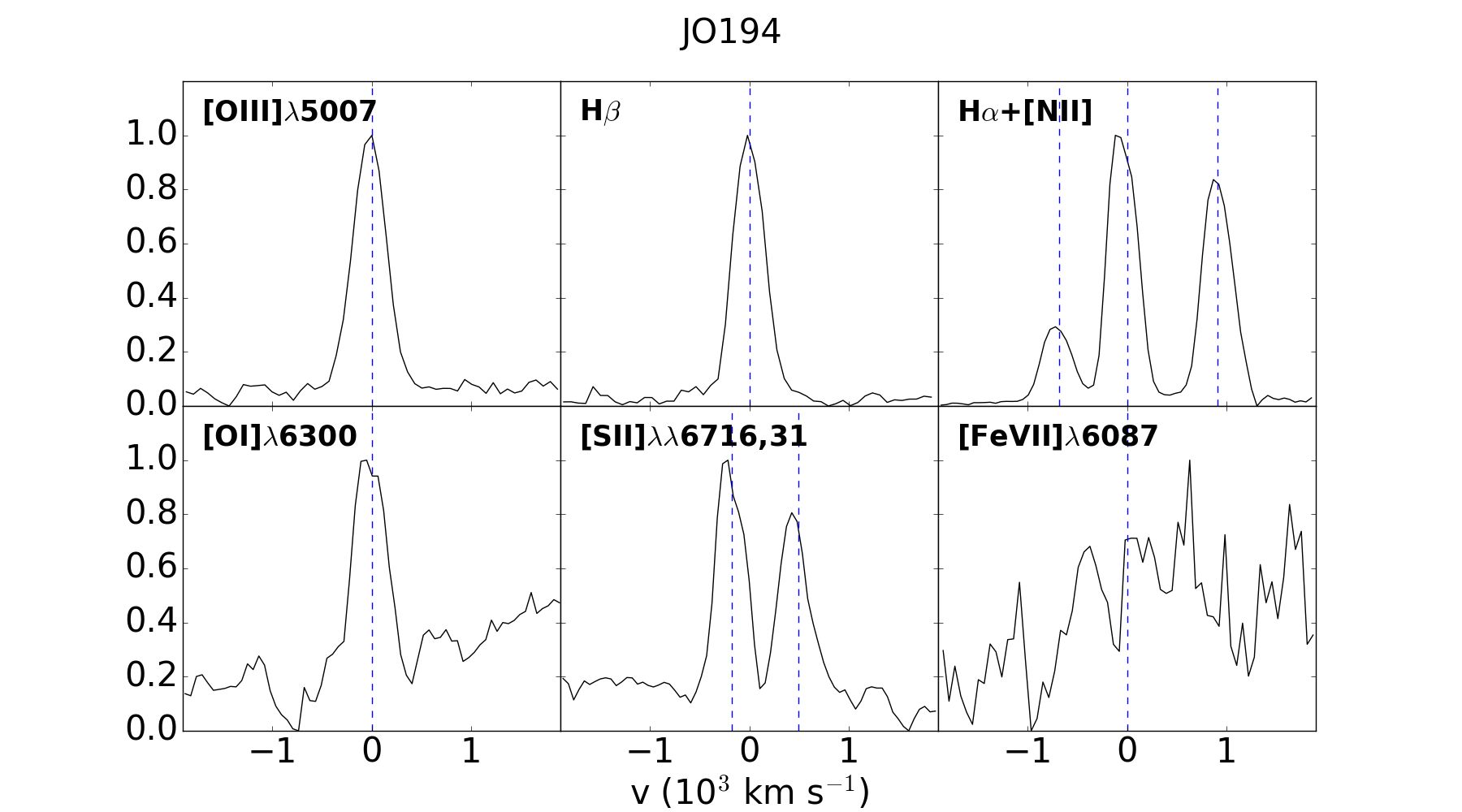} \\
	\end{tabular} 
	\contcaption{}
\end{figure*}
	
\begin{figure*}
	\begin{tabular}{cc}		
	\includegraphics[width=.5\linewidth]{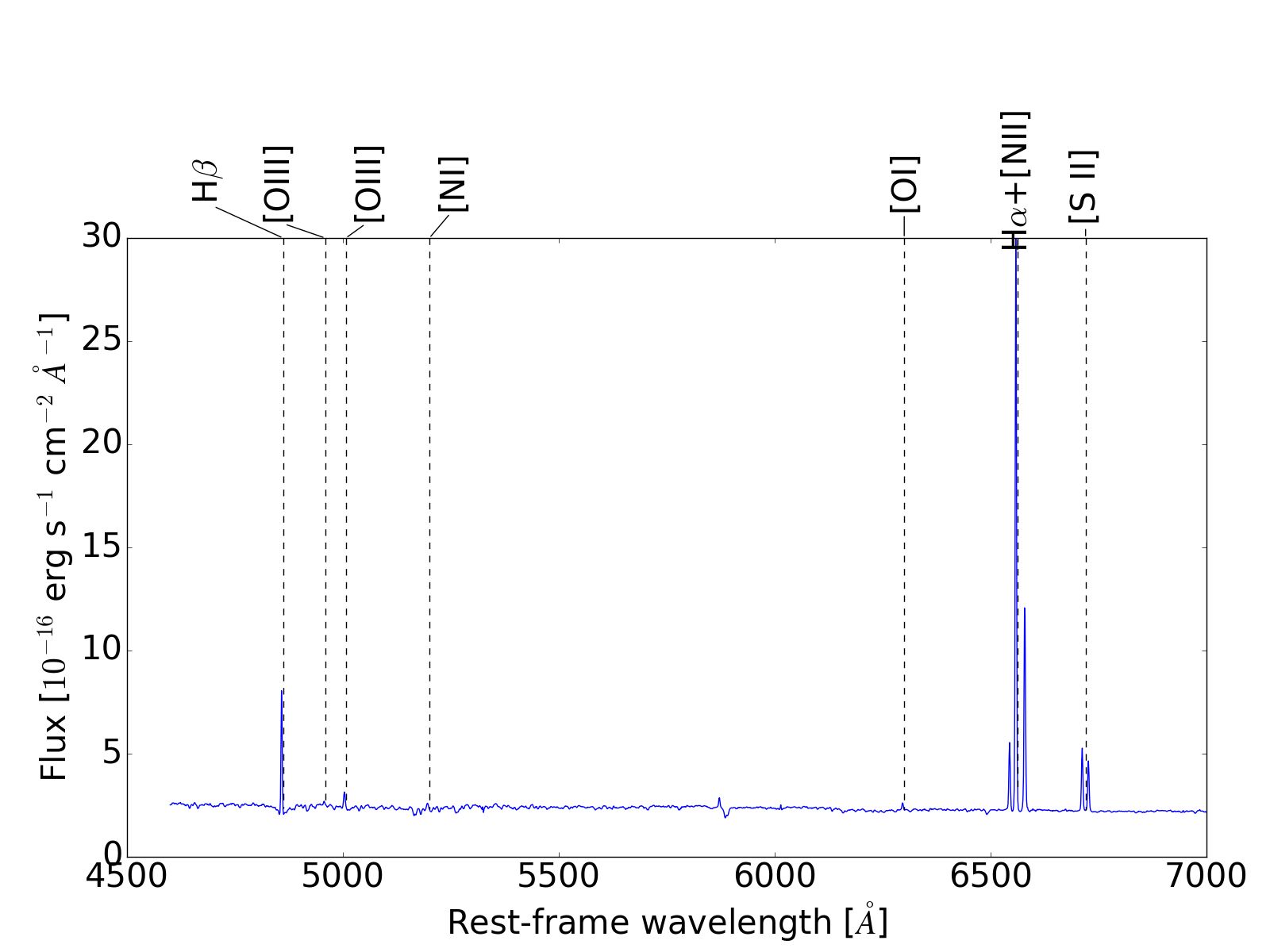}	&
	\includegraphics[width=.5\linewidth]{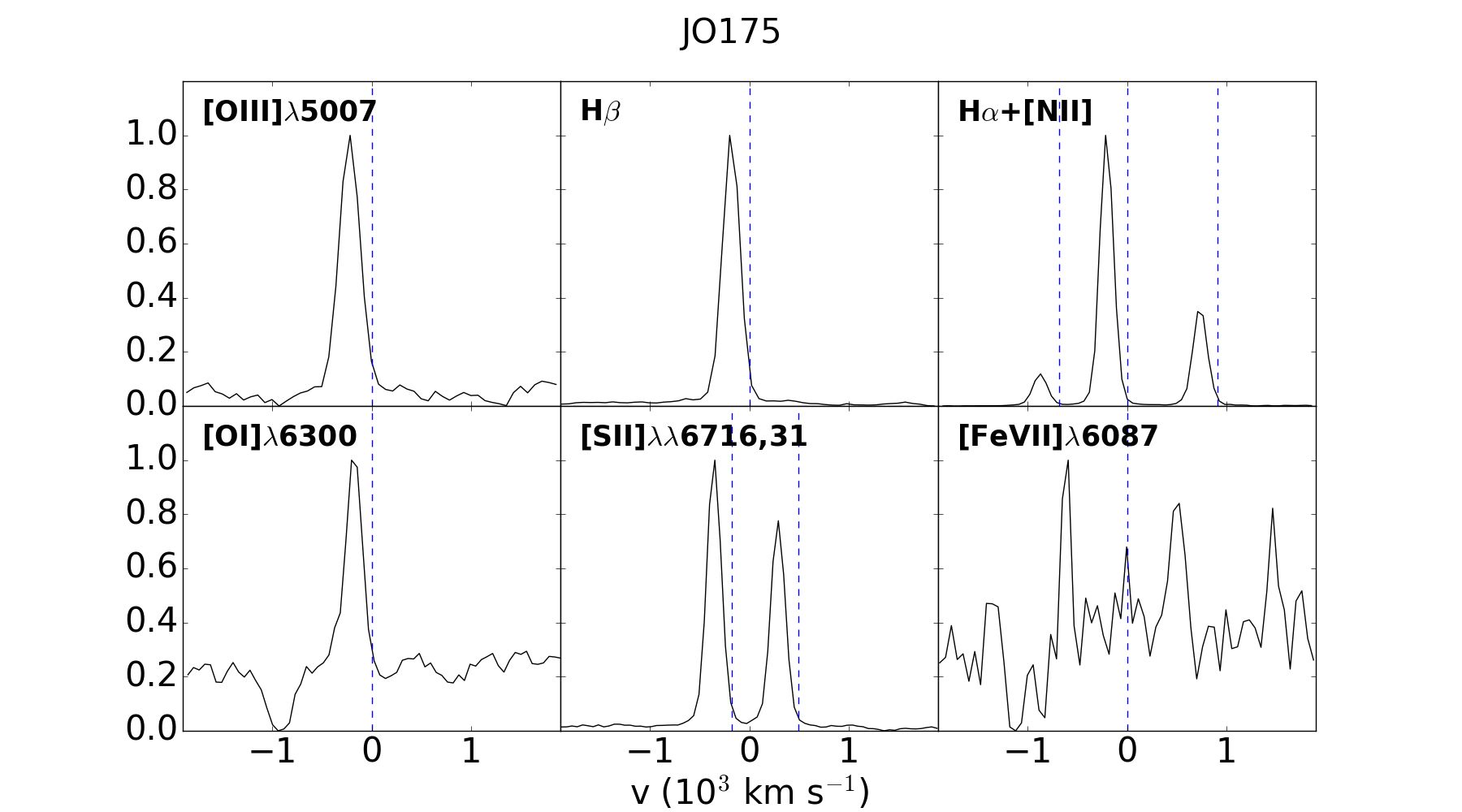} \\
	\end{tabular} 
	\contcaption{}
\end{figure*}

\section{Galaxy sample, observations and data analysis}
\label{sec:Data}

In this paper we analyze the seven galaxies in P17b (Table~\ref{tab:props} and Fig.~\ref{fig:VRI}), which are all characterized by tails of ionized gas at least as long as the stellar galaxy diameter. These galaxies represent extreme cases where the cluster environment strongly acts on the gas and possibly on the AGN. Spectra and individual emission lines for the central spaxel of each galaxy, selected as discussed later, are displayed in Fig.~\ref{fig:spectra}. 
In the following description,  a reference is given in square brackets after the galaxy name  for those galaxies studied individually in a GASP paper. 

\noindent {\bf JO201}, or Kaz 364 [P17a, \citet{2017ApJ...844...49B},\citet{2018MNRAS.479.4126G}]
was also classified  by \citet{2009ApJ...707.1691A} as an AGN based on XMM observations.
Two components are  present in the nuclear emission lines
(P17b, see also 
Fig.~\ref{fig:spectra}): a narrow, stronger component, and a broader one, slightly blueshifted.

\noindent {\bf JO204} [\citet{2017ApJ...846...27G}] was included by \citet{2016ApJ...832...67N} in a sample of 71 double-peaked AGN selected from the SDSS and classified as an AGN with an outflow. At least two components are visible in the MUSE spectra; an extended emission with AGN-like line ratios is detected up to $\sim 20$ kpc from the nucleus, see below Sect.~\ref{sect:ExtNucleus}.

\noindent {\bf JO135}  
Complex emission line profiles are observed in the nucleus, with a strong redshifted wing (Fig.~\ref{fig:spectra}). 
AGN-like emission is present up to $\sim$ 10 kpc (P17b), see Sect.~\ref{sect:ExtNucleus}.

\noindent {\bf JW100}, or IC 5337, was  classified as an AGN by \citet{2008ApJ...682..155W}, based on X--ray Chandra observations. It was classified as a head-tail radio source by
\citet{2013MNRAS.436L..84G}, who detected radio emission in Very Large Array radio measurements at 1.4 and 4.8 GHz: the peak of the radio emission coincides with the MUSE center.
Double-peaked profiles are detected in the region around the nucleus (P17b).

\noindent{\bf JO175}, {\bf JO206}  [P17a], {\bf JO194}: in these galaxies emission lines appear as  single-component Gaussians.

We refer to P17a for a detailed description of the GASP survey,  data and adopted reduction techniques.
Observations were obtained with the MUSE spectrograph in wide-field mode with natural seeing \citep{2010SPIE.7735E..08B}. One or two MUSE pointings per galaxy, each with a 2700sec exposure and covering a 1'x1' field of view, are sampled with 0.2"x0.2" pixels over the spectral range 4800-9300 \AA $\,$ with a spectral resolution FHWM $\sim 2.6$ \AA. Data were taken under clear dark sky conditions, with $<1"$ seeing (Table~\ref{tab:props}).  
We remind the reader that the fitting of emission lines in GASP was done  using \texttt{KubeViz}  \citep{2016MNRAS.455.2028F}. Velocity and velocity dispersion were derived from the fit of the lineset consisting of  H$\alpha$ and the [NII]$\lambda\lambda$6548,6583 doublet, and used for all other lines; as necessary, one or two Gaussian components were adopted.
Emission line velocity dispersions ($\sigma_{obs}$) were corrected for the instrumental component: $\sigma = \sqrt{\sigma_{obs}^{2}-\sigma_{inst}^{2}}$; $\sigma_{inst}$ was derived at each wavelength using a third order polynomial fit of the MUSE  resolution curve \citep{2014MNRAS.445.4335F}.
In the following, we will use the data cube average filtered with a 5x5 pixel kernel in the spatial direction, unless otherwise stated, having subtracted the stellar underlying spectrum fitted with the \texttt{SINOPSIS} code \citep{2017ApJ...848..132F}. The stellar kinematics is derived with the \texttt{pPXF} code \citep{2004PASP..116..138C}  using stellar population templates from \citet{2010MNRAS.404.1639V}, see P17a for details. The galaxy center was defined as the centroid of the  continuum map obtained by the  \texttt{KubeViz} best fit model in the H$\alpha$ region.

\begin{figure*}
	\begin{tabular}{cc}
		JO135 & JO201\\
		\includegraphics[width=.5\linewidth]{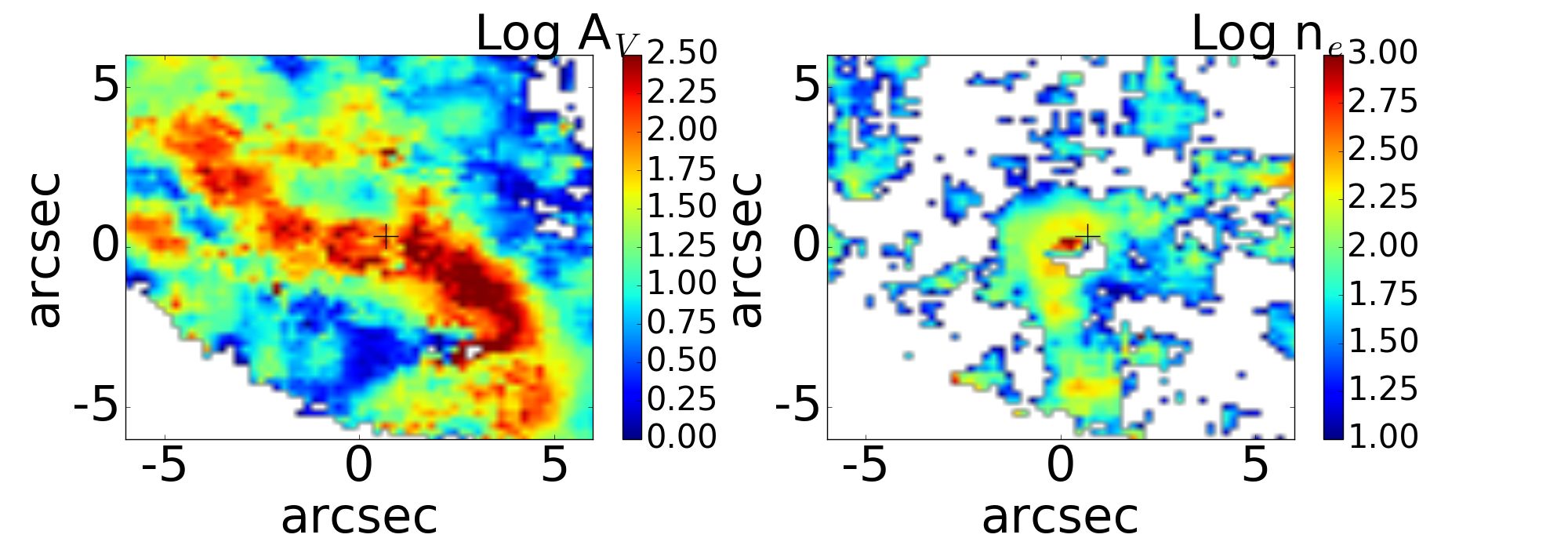} &
		\includegraphics[width=.5\linewidth]{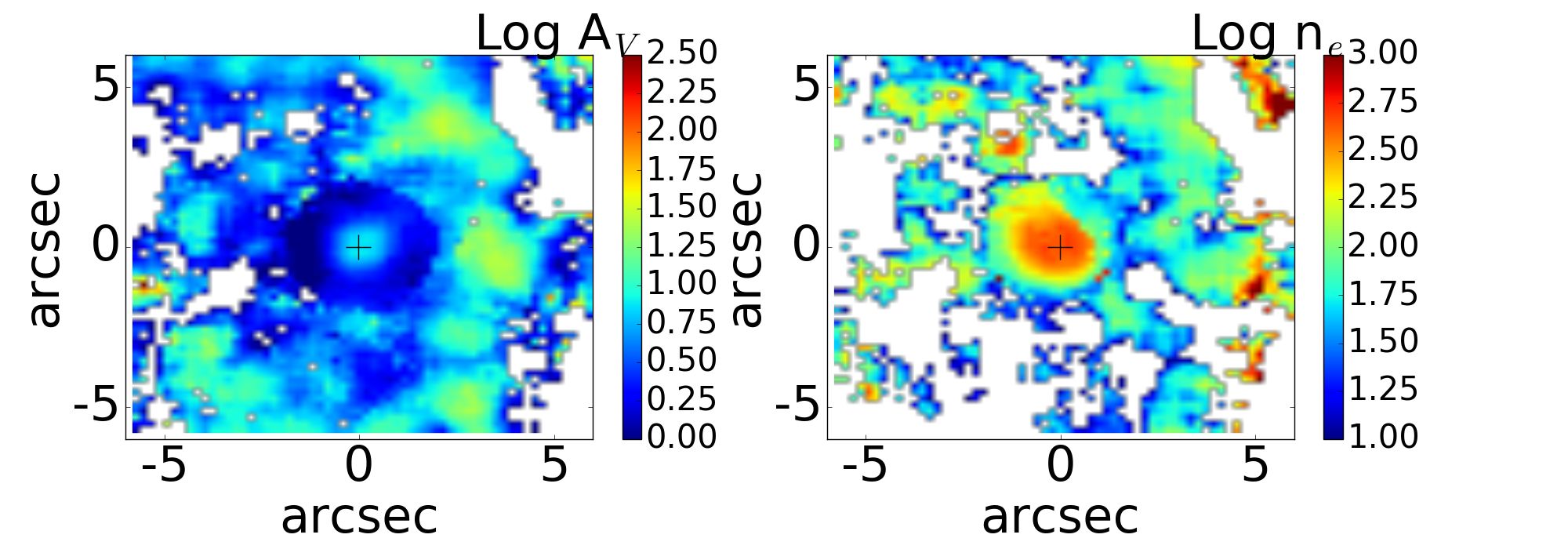} 
		\\		
		JO204 & JO206\\
		\includegraphics[width=.5\linewidth]{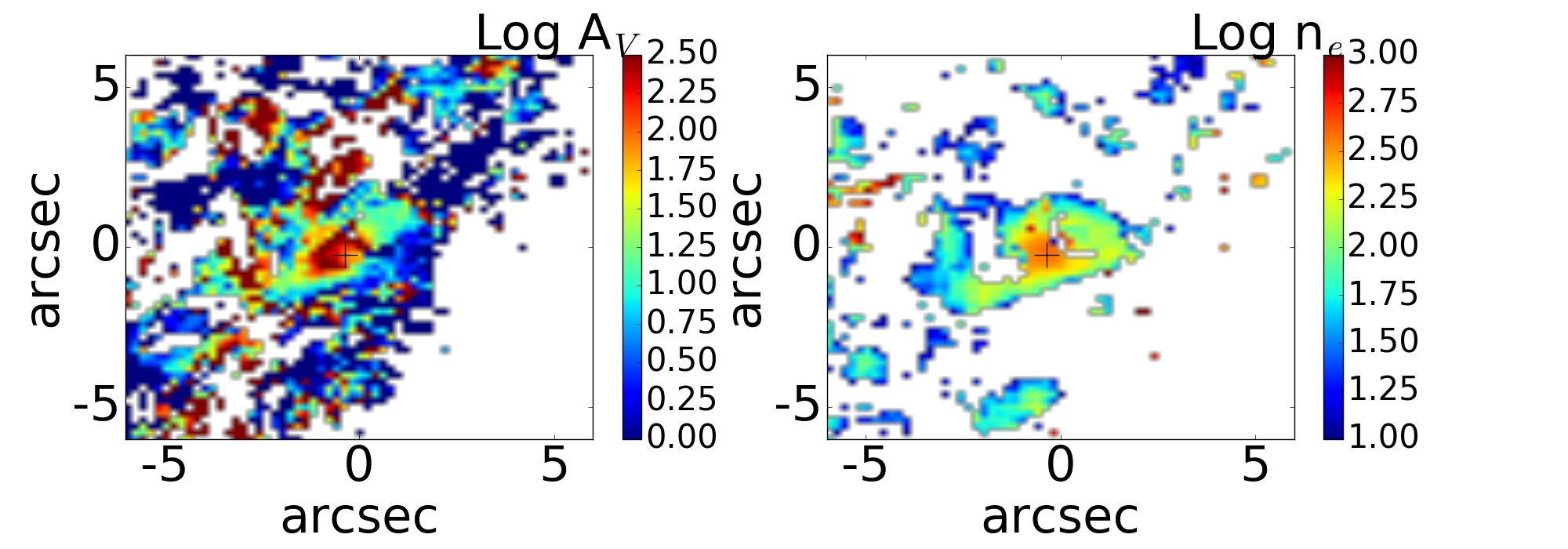} &
		\includegraphics[width=.5\linewidth]{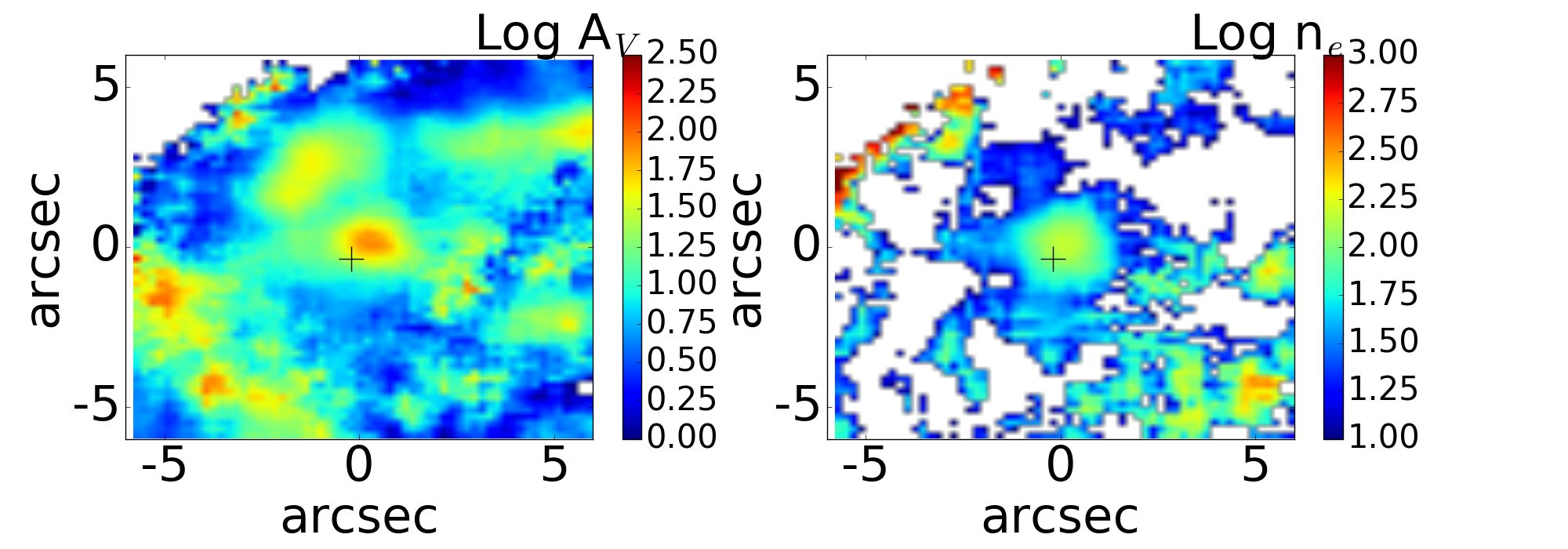} \\
		JW100 & JO175\\
		\includegraphics[width=.5\linewidth]{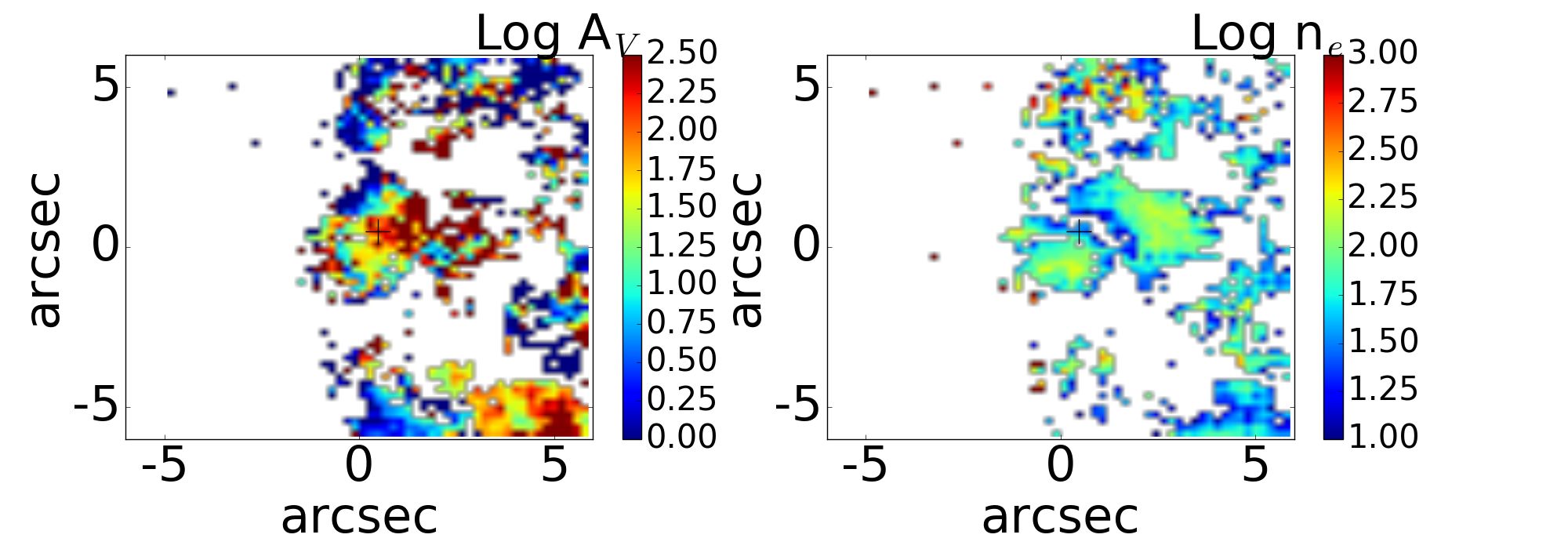} &
		\includegraphics[width=.5\linewidth]{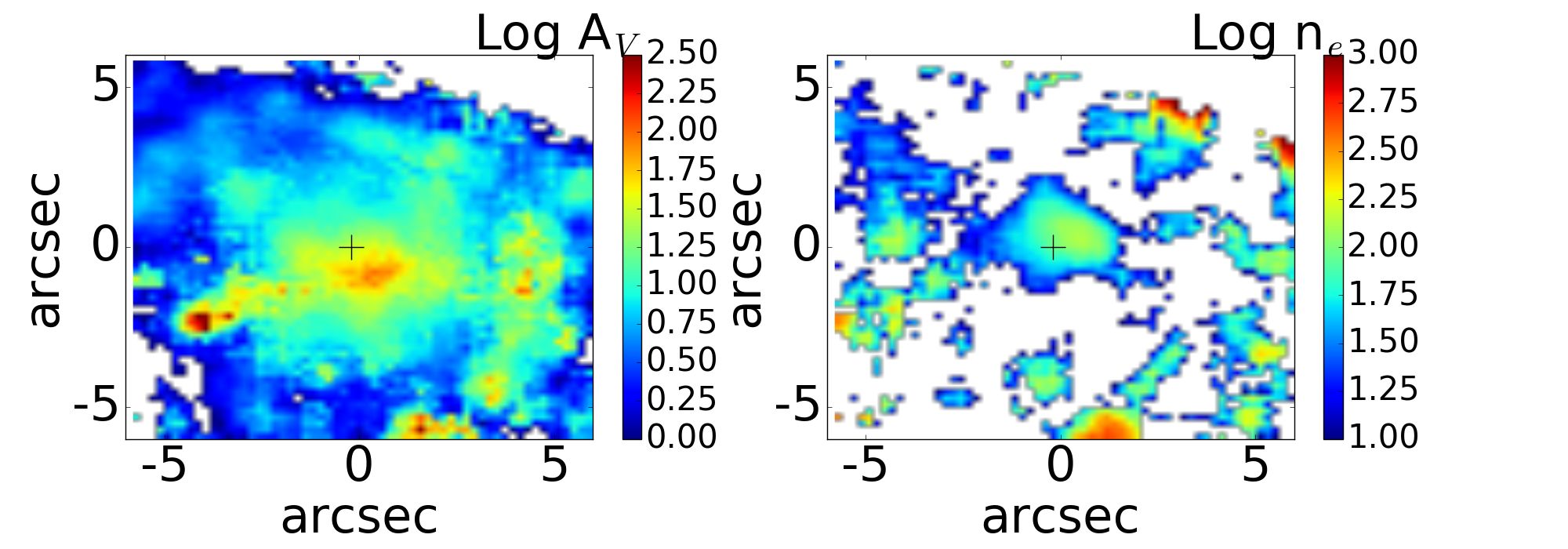} \\
		JO194 \\
		\includegraphics[width=.5\linewidth]{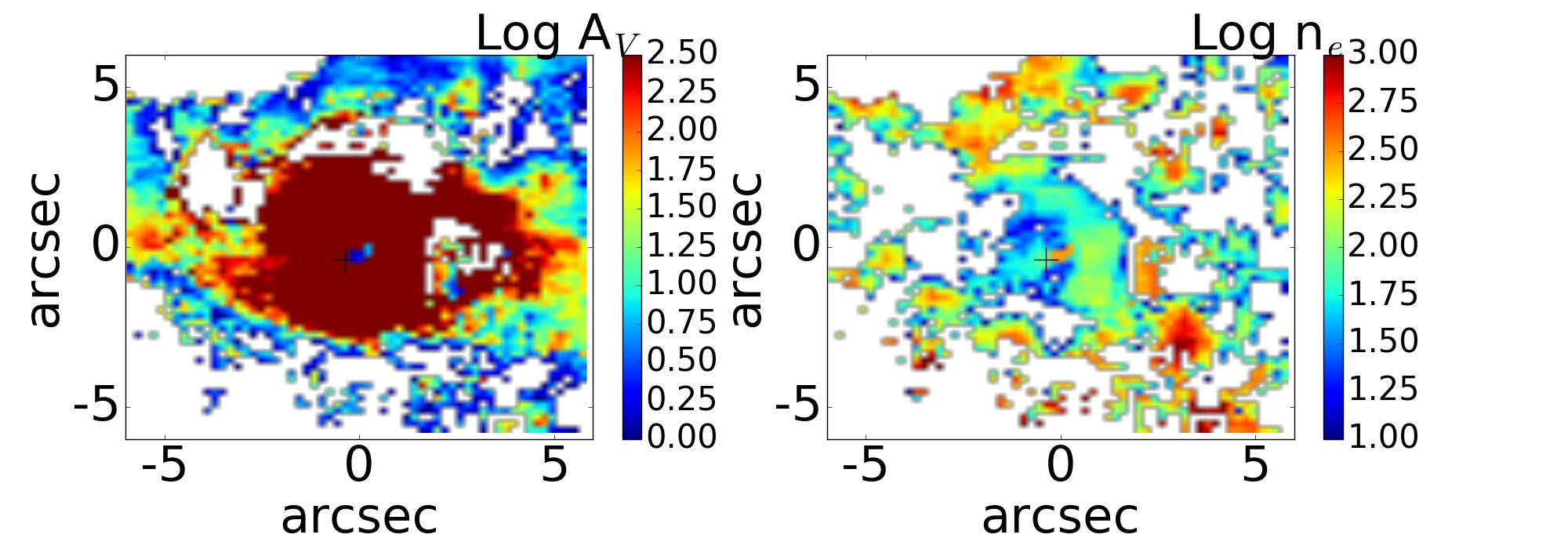} \\

	\end{tabular} 
	\caption{For each galaxy the plots show the spatial distribution in the central 10\arcsec x 10\arcsec of {\em left:} extinction ($A_v$), {\em right:} $\log{n_e}$. For all galaxies, 1 arcsec is $\sim 1$ kpc (see Table~\ref{tab:props}).\label{fig:dpars}}
\end{figure*}

All the line fluxes, selected to have a signal to noise ratio SN >3, were corrected for dust extinction using the Balmer decrement as in P17a. Fig.~\ref{fig:dpars} displays the extinction map 
($A_v$) in each galaxy, as well as the electron density ($n_e$): this was derived using the relations in \citet{2014A&A...561A..10P},  as in P17a. 
We detect increased values of the extinction ($A_v$ > 2) in the central spaxels of JO135, JO204, JO206  and JW100. In JO201 the extinction is low in the nucleus ($A_v$ < 1). In JO194 the extinction is high ($A_v > 2.5$) in an extended region of radius $\sim 3\arcsec$ around the central spaxel.  JO201, JO204 and JO135  show a steep increase in density ($n_e$ > $10^{2.5}$ cm$^{-3}$) in the nucleus: lower densities are measured in JO206 and JW100 ($n_e$ $\sim$ $10^2$ cm$^{-3}$), and JO194, JO175 ($n_e$ < $10^2$ cm$^{-3}$) .

\begin{figure*}
	\begin{tabular}{c}
		JO135 \\
		\includegraphics[width=.95\linewidth]{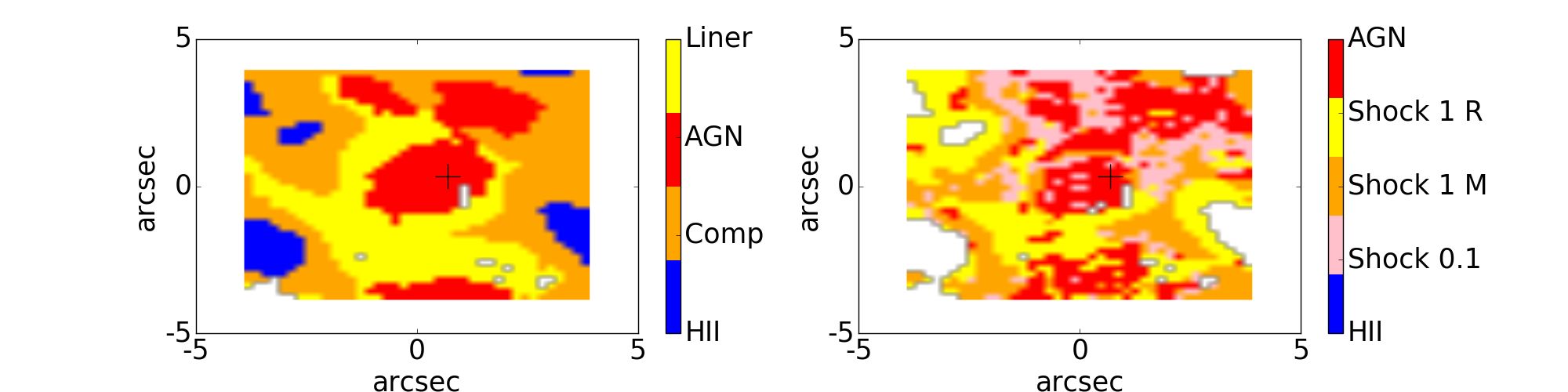}\\
		JO201\\
		\includegraphics[width=.95\linewidth]{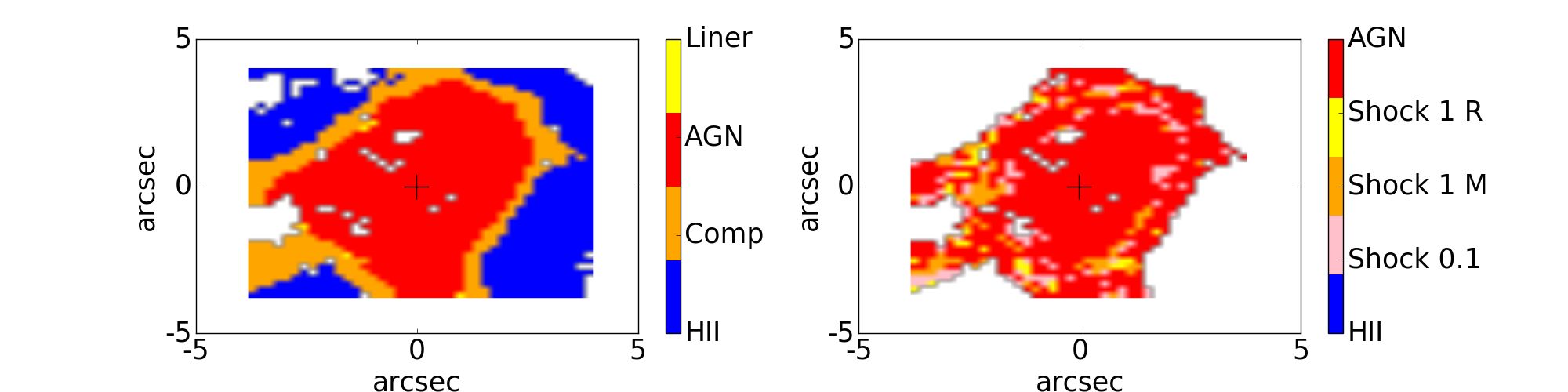}\\
		JO204\\
		\includegraphics[width=.95\linewidth]{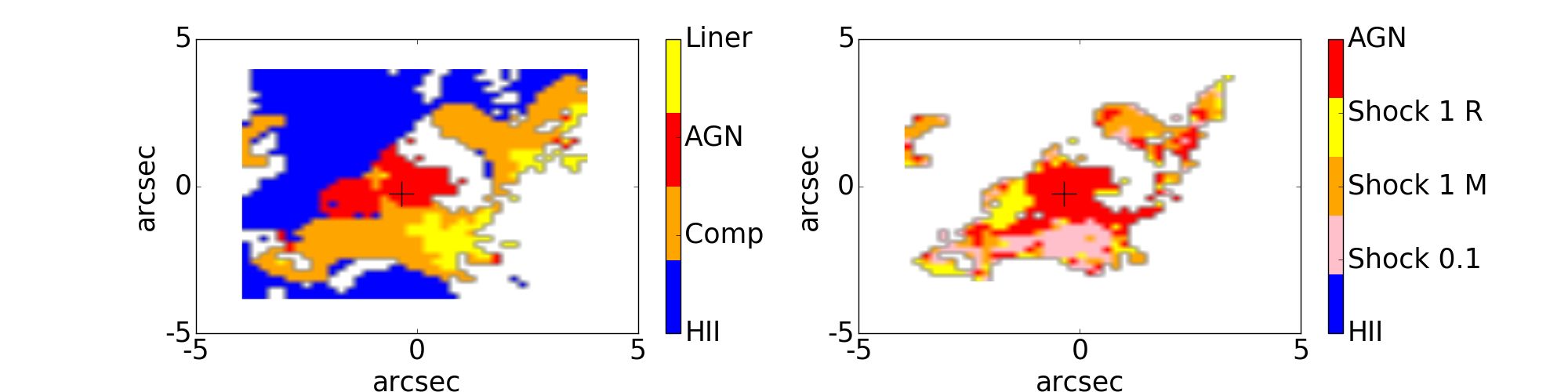}\\
		JO206 \\
		\includegraphics[width=.95\linewidth]{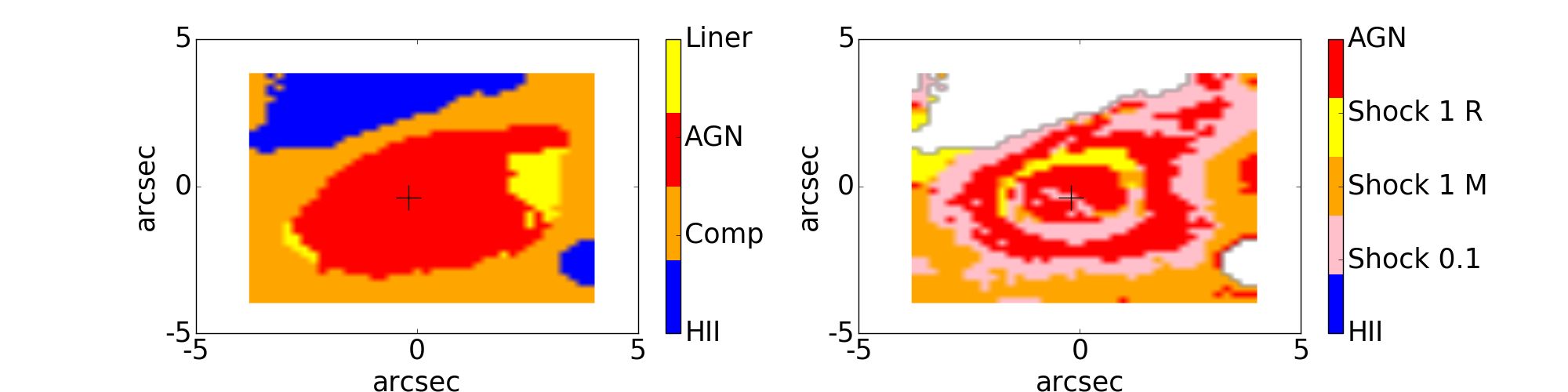}\\
	\end{tabular} 
	\caption{Color coded maps in the a region of 100x100 spaxels around the nucleus. {\em Left}: classification from P17b (HII, composite, AGN, LINER). {\em Right:} NB models: HII; shock with $n=0.1$ cm$^{-3}$, solar abundances; shock with $n=1$ cm$^{-3}$, solar abundances (M); shock with $n=1$ cm$^{-3}$, 2x solar abundances (R); AGN. Spaxels classified as HII in P17b were not fitted with NB. \label{fig:mapsKV2NB}}
\end{figure*}

\begin{figure*}
	\begin{tabular}{c}	
		JW100 \\ \includegraphics[width=.95\linewidth]{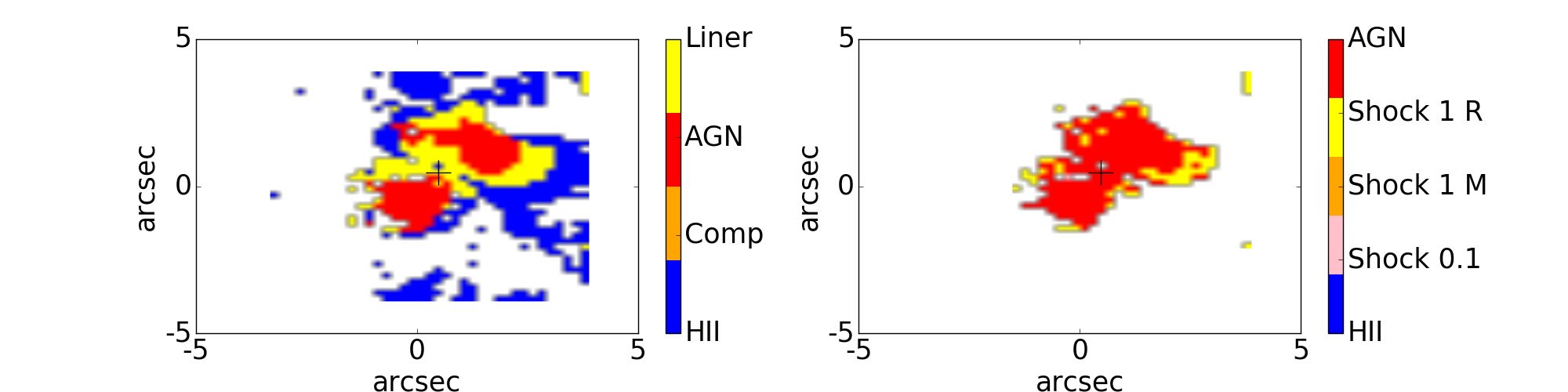}\\
		JO194 \\	\includegraphics[width=.95\linewidth]{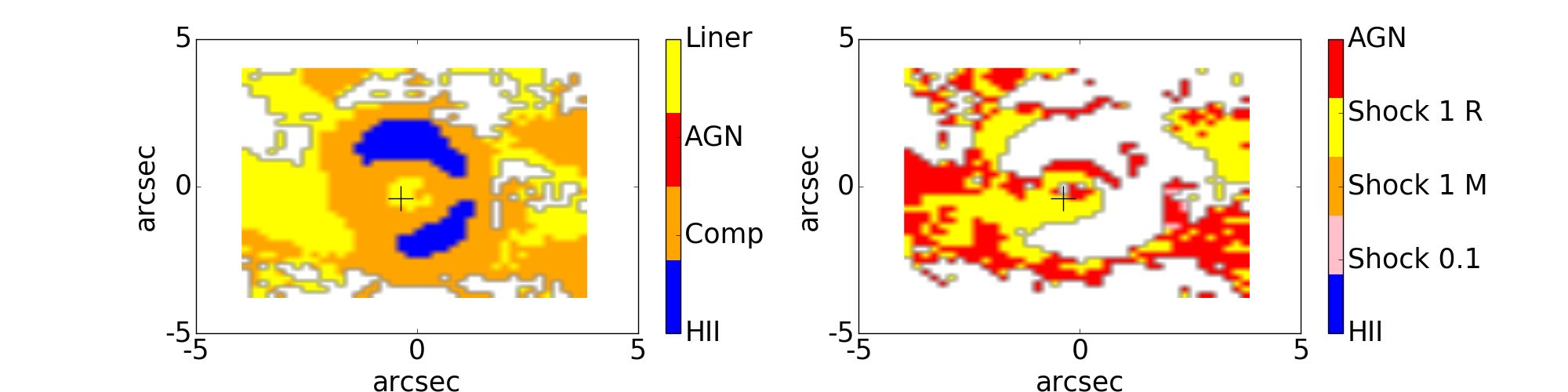}\\
		JO175 \\	\includegraphics[width=.95\linewidth]{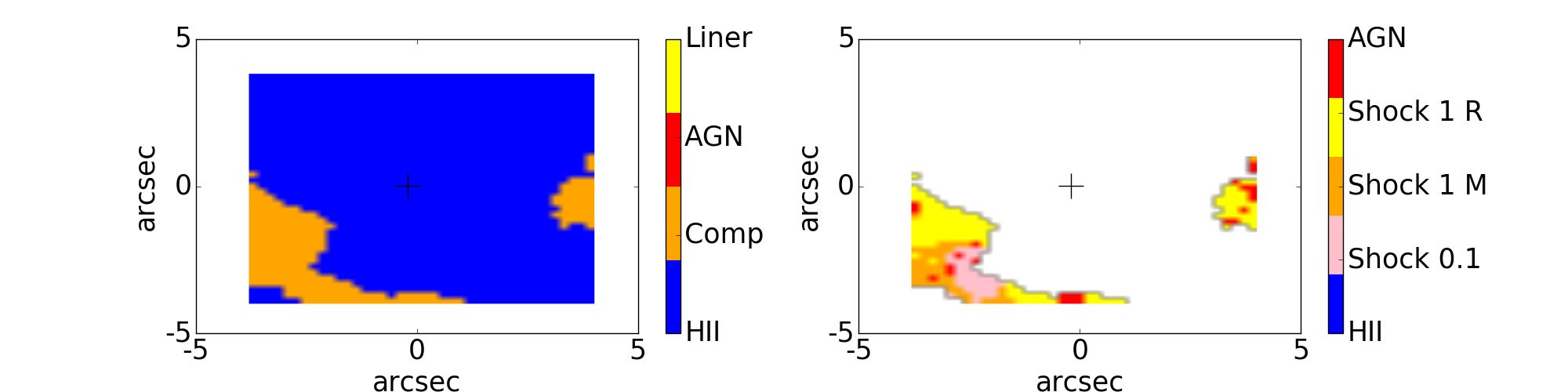}\\	
	\end{tabular} 
	\contcaption{}
\end{figure*}

\begin{figure*}

\begin{tabular}{c}
JO135: observed vs. AGN models \\
\includegraphics[width=.8\linewidth]{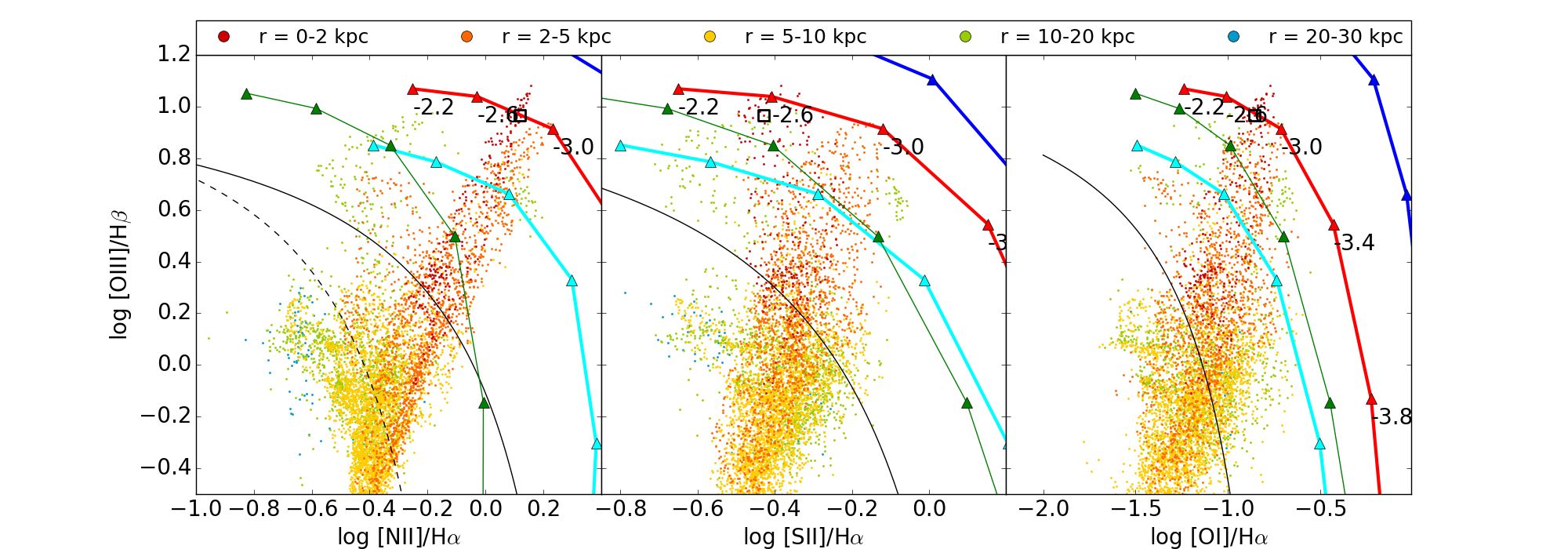}\\
JO135: observed vs. shock models \\
\includegraphics[width=.8\linewidth]{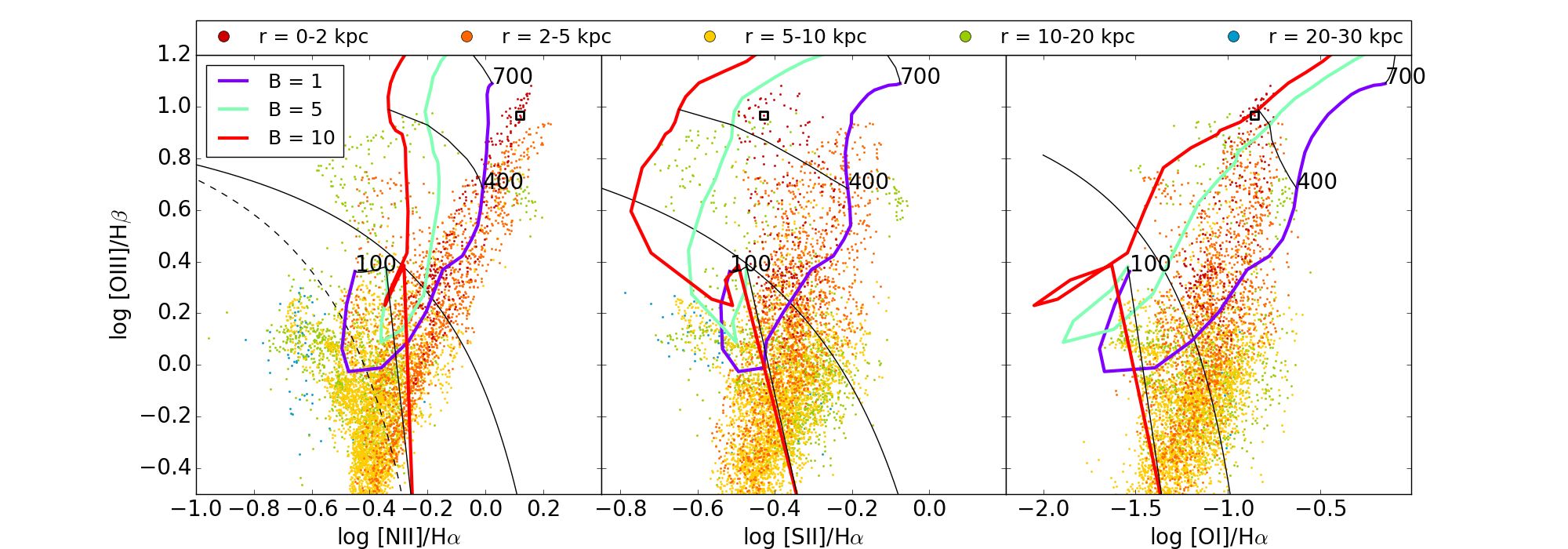}\\

JO201: observed vs. AGN models \\
\includegraphics[width=.8\linewidth]{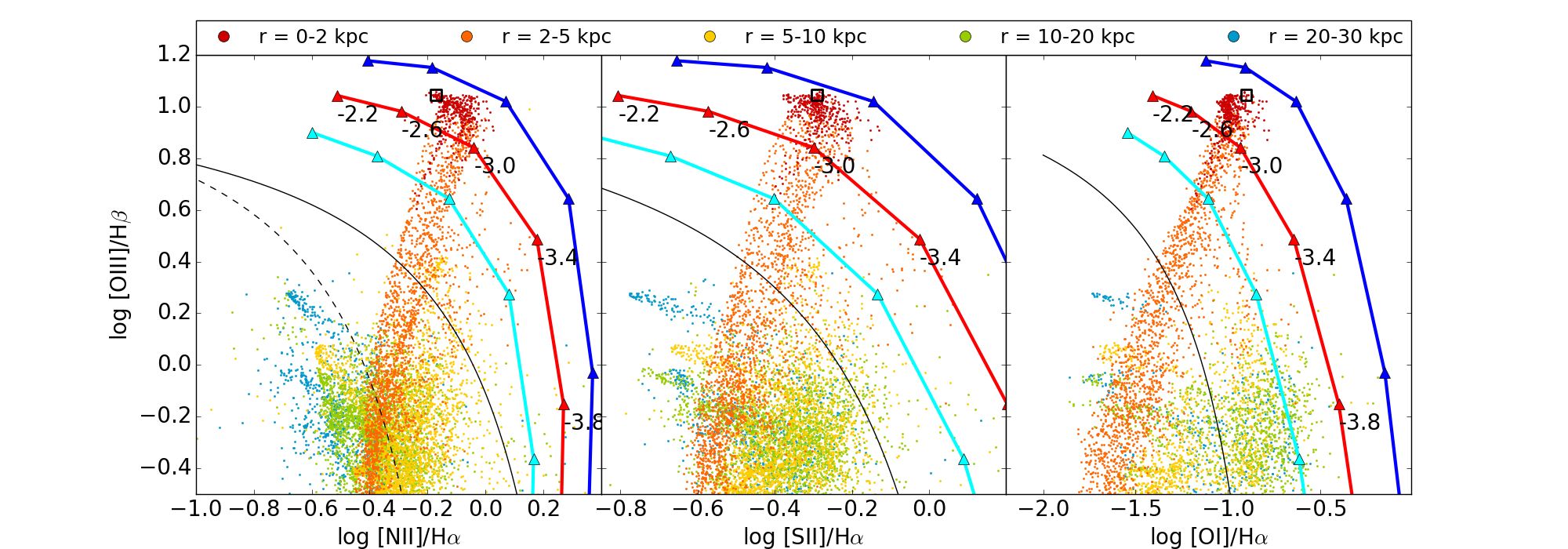}\\
JO201: observed vs. shock models \\
\includegraphics[width=.8\linewidth]{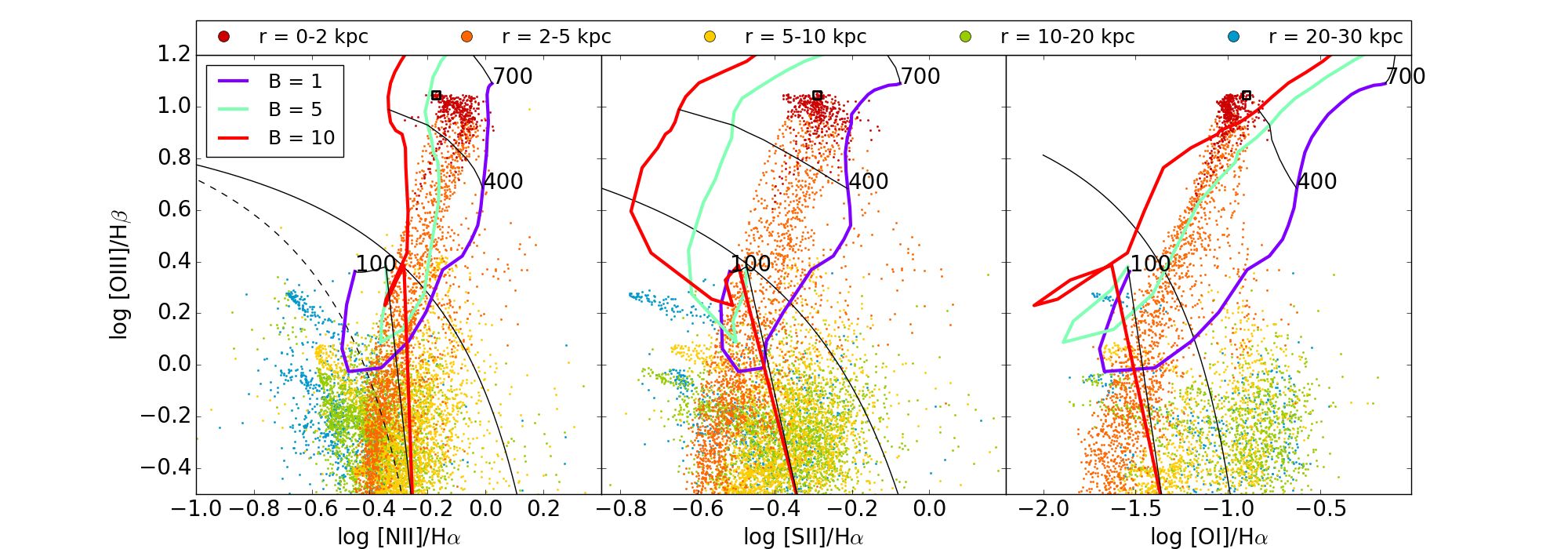}\\

\end{tabular} 
\caption{Observed emission line ratios color coded with the  projected distance from the center; the empty square displays the value measured in the central spaxel. The black solid and dashed ([NII]/H$\alpha$ panel) curves indicate the empirical SF/Composite/AGN classification by \citet{2006MNRAS.372..961K}.	
	For each galaxy, overlaid are best-fit AGN (not for JO175) and shock models.  {\bf AGN models} -- The red lines display  models for different values of $\log U$, adopting the best-fit values of  $\log P/k$, $12 + \log O/H$  and  $E_{\rm peak}$ in the central spaxels; models with an offset $\pm$ 0.25  in  $E_{\rm peak}$ are displayed in cyan and blue respectively. For JO204 and JO135, the green line shows AGN models in the EENLR. {\bf Shock models} -- The lines display models for $v_{\rm sh}$ = 100-700 km s$^{-1}$, pre-shock density $n=0.1$ cm$^{-3}$, solar abundances (JO135, JO201,  JO206);  $n=1$ cm$^{-3}$, 2x solar abundances (JO204, JW100, JO175, JO194), and magnetic field $B$ as in the legend. 	\label{fig:models}}
\end{figure*}

\begin{figure*}
	\begin{tabular}{c}
JO204: observed vs. AGN models \\
		\includegraphics[width=.8\linewidth]{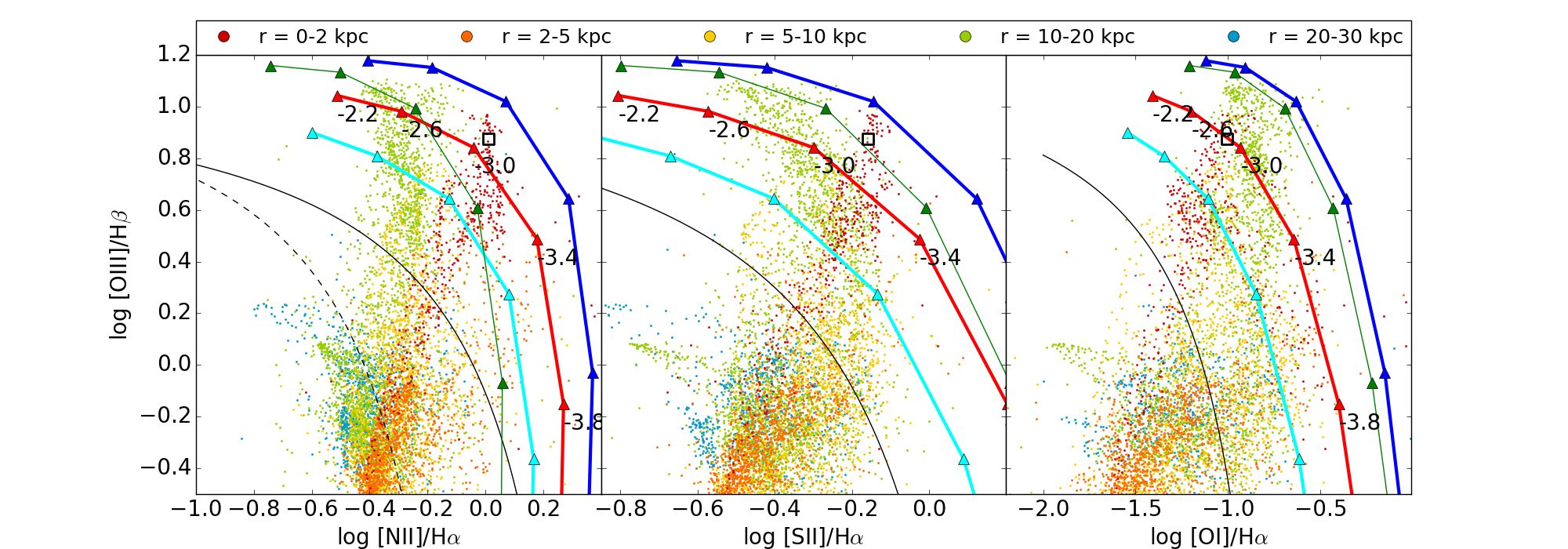}\\
JO204: observed vs. shock models \\
		\includegraphics[width=.8\linewidth]{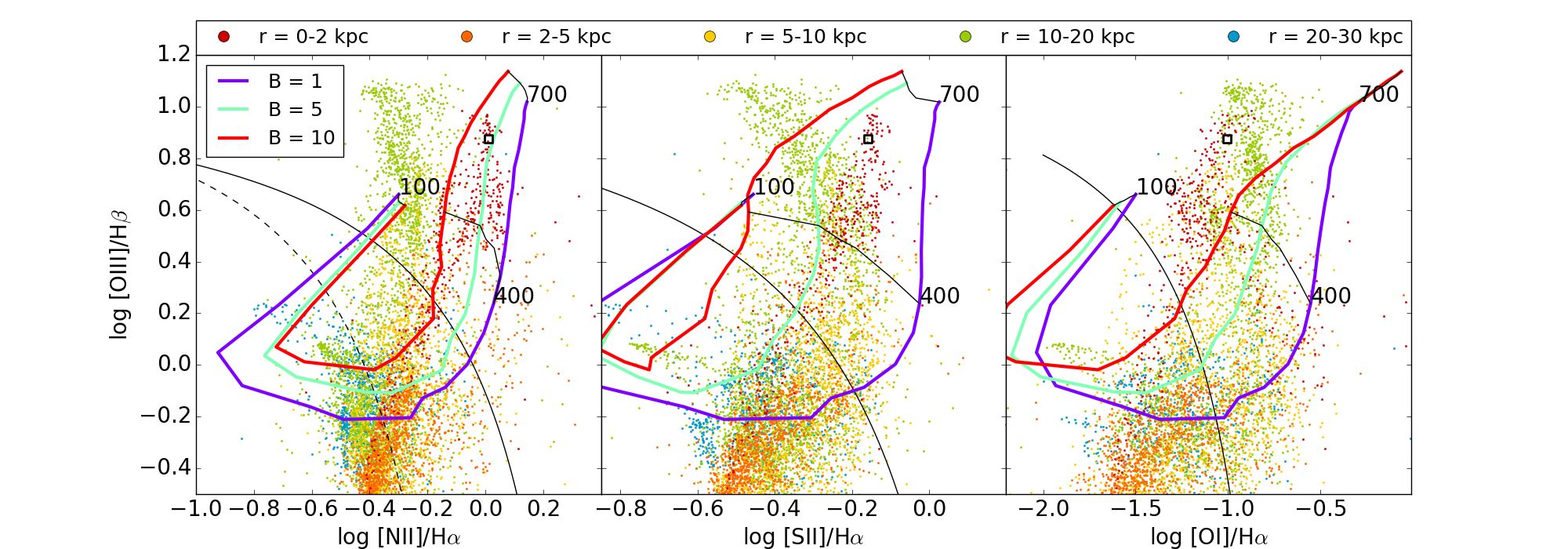}\\
%		(c) JO206 \\	
JO206: observed vs. AGN models \\	
		\includegraphics[width=.8\linewidth]{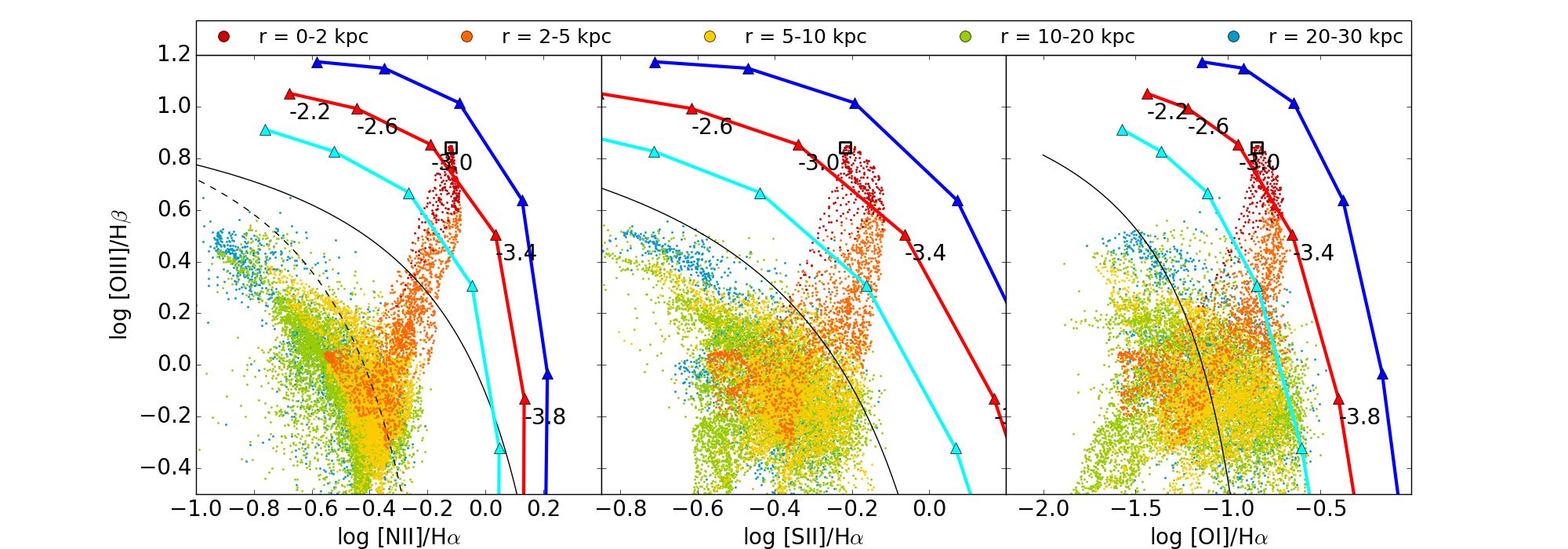}\\
JO206: observed vs. shock models \\
		\includegraphics[width=.8\linewidth]{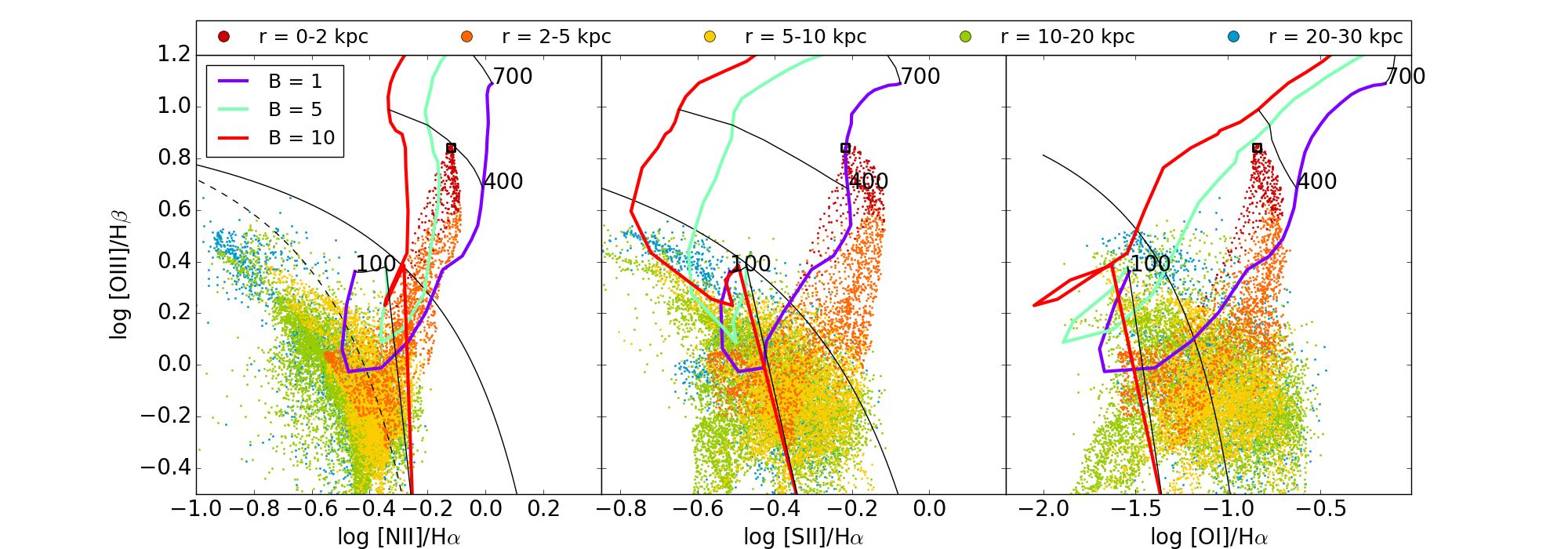}\\
	\end{tabular} 
	\contcaption{}
\end{figure*}

\begin{figure*}
	\begin{tabular}{c}
JW100: observed vs. AGN models \\
		\includegraphics[width=.8\linewidth]{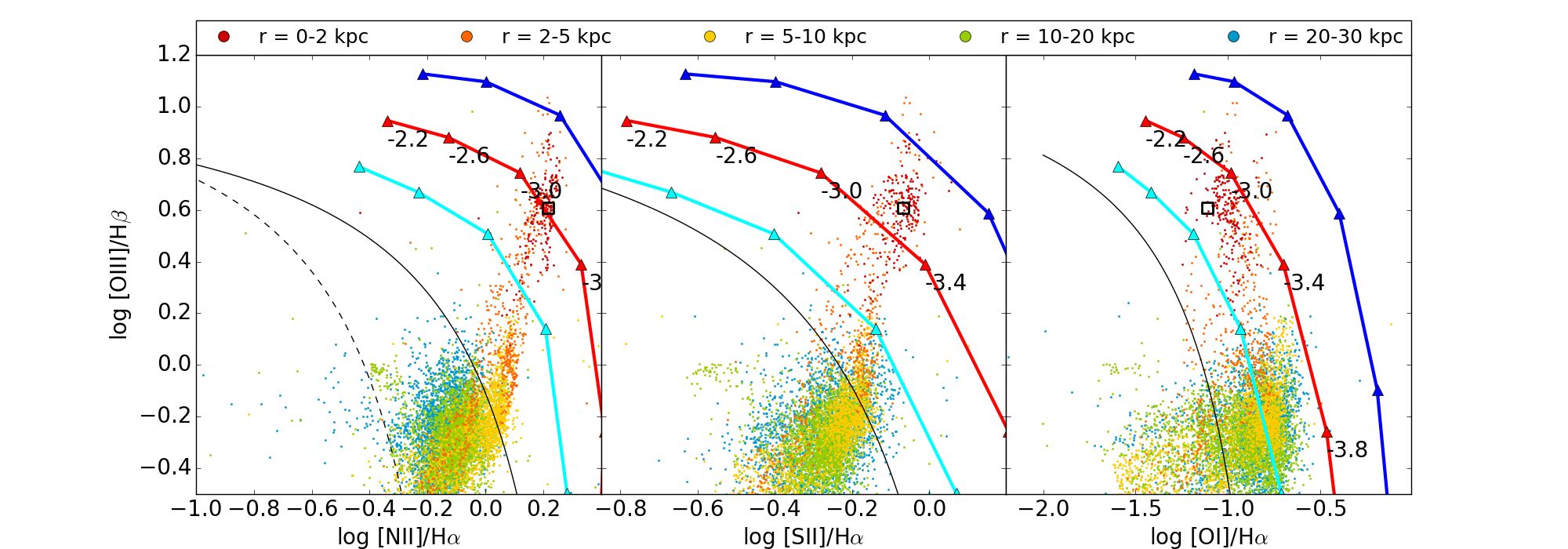}\\
JW100: observed vs. shock models \\
		\includegraphics[width=.8\linewidth]{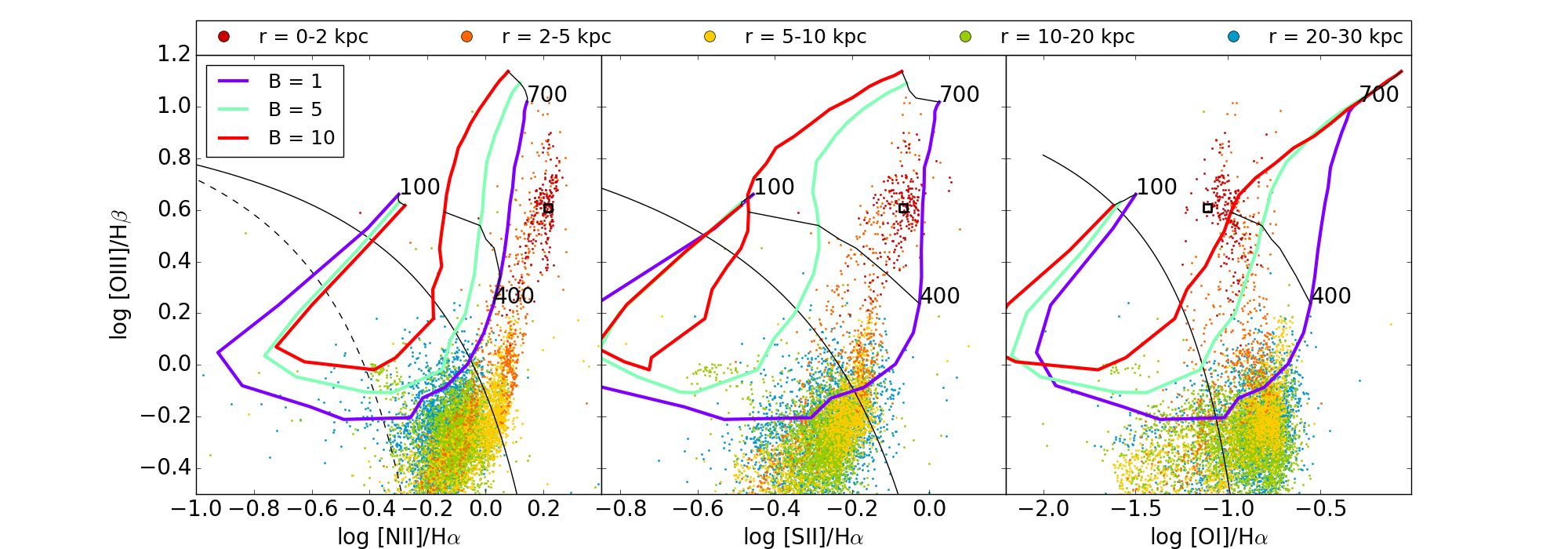}\\
JO194: observed vs. AGN models \\
		\includegraphics[width=.8\linewidth]{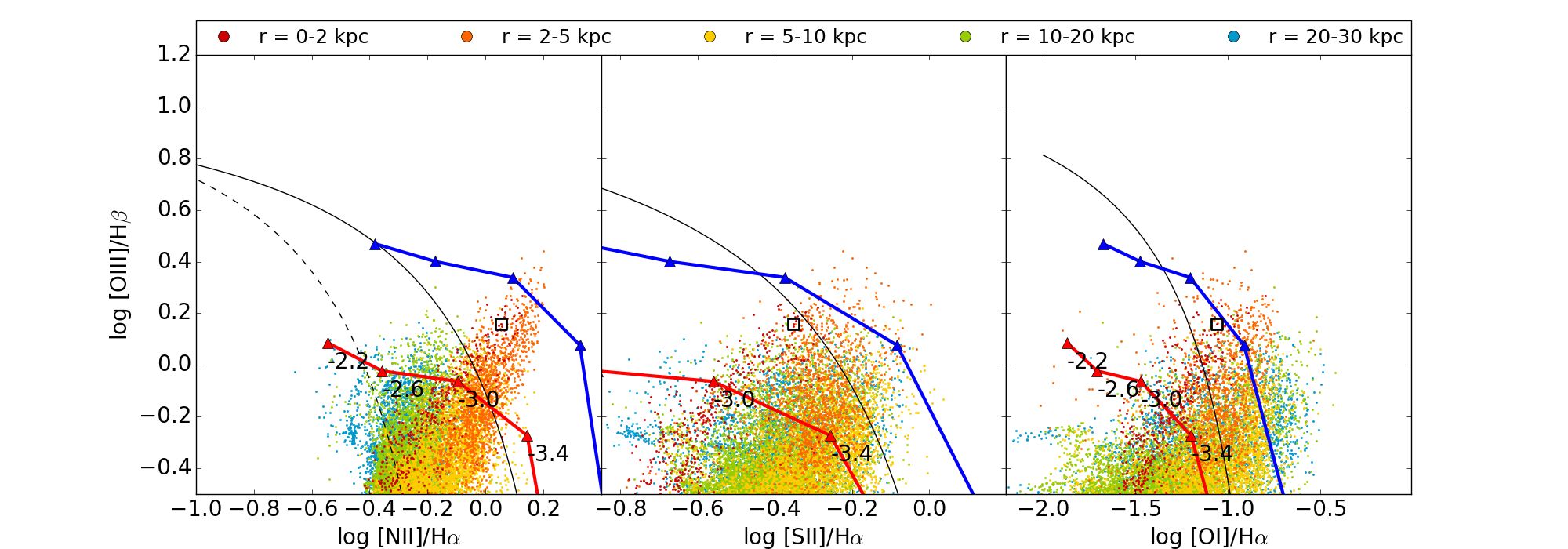}\\
JO194: observed vs.shock models \\
\includegraphics[width=.8\linewidth]{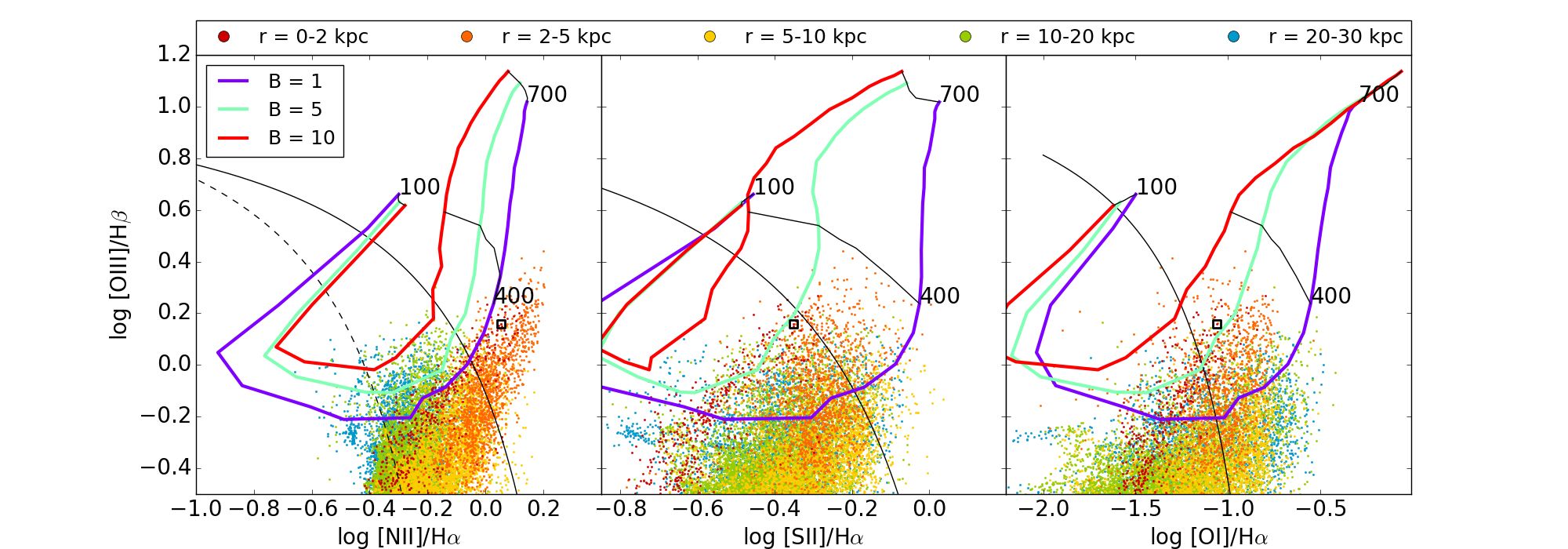}\\
	\end{tabular} 
	\contcaption{}
\end{figure*}

\begin{figure*}
	\begin{tabular}{c}%		(f) JO175\\
JO175: observed vs. shock models \\		
\includegraphics[width=.8\linewidth]{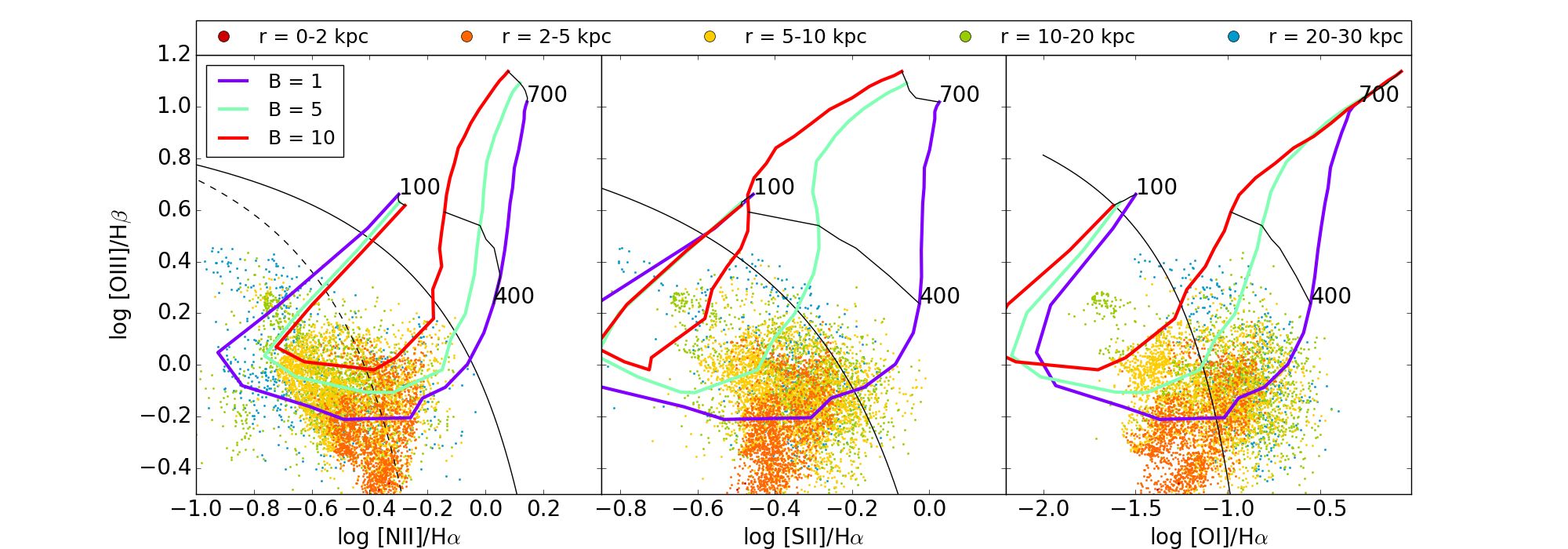}\\
\end{tabular} 
\contcaption{}
\end{figure*}

%============================================================

\section{Ionization mechanisms}
\label{sect:analysis}

The left panels of Fig.~\ref{fig:mapsKV2NB} present the classification in HII-regions, Composite, AGN and Liners based on [NII]$\lambda$6583/H$\alpha$ vs. [OIII]$\lambda$5007/H$\beta$ as in P17b, whose main conclusions are summarized here.
In some cases (JW100, JO135\footnote{Compared to P17b, we improved the fitting of the lines of the central spaxels of JO135.}, JO201, JO204) one Gaussian was not enough to fit the emission line profiles and a double-Gaussian fit was adopted: in these cases, the classification displayed in Fig.~\ref{fig:mapsKV2NB} refers to the narrow component.

Based on the P17b analysis,
JO201, JO204, JO206, JO135 and JW100 present AGN-like line ratios in the inner kpcs.
None of them shows broad (> 5000 km s$^{-1}$), permitted lines typical of the Broad Line Region in AGN; the observed emission lines are therefore produced in the Narrow Line Region (NLR). Extended AGN-like emission over several kpcs is observed in JO204 and JO135, and it can be attributed to anisotropic ionization from the AGN (the so-called ionization cones).
In JO175 the emission is mostly due to star formation, while in 
JO194, composite  line ratios are detected throughout the galaxy.

We define the  size of the Narrow Line Region in the AGN candidates  as in \citet{2017ApJ...837...91B}, that is  weighting the projected distances from central spaxel on the [OIII]$\lambda$5007 fluxes, over the spaxels 
classified as AGN in P17b: 
\begin{equation}
r_{\rm NLR}=\sum_{\rm{r<5 Kpc}} r f_{\rm[OIII]}(r) /\sum_{\rm{r<5 Kpc}} f_{\rm[OIII]}(r),
\end{equation} 
$r$ being the distance from the central spaxel and  $f_{\rm[OIII]}$ the line flux at that distance.
The NLR size so defined is of the order of 1 kpc for all galaxies: as the typical seeing was $\sim$ 0.8-1 arcsec, the NLR is unresolved. To have an estimate of the maximum extension of the AGN-ionized region, we also computed the 95\% percentile of the distances for the AGN spaxels ($r_{\rm AGN}$ in Table~\ref{tab:props}), keeping in mind that these values may be biased by spaxels with fainter [OIII] fluxes, where the measurement uncertainties may produce a wrong classification. The AGN emission can be therefore described as the sum of a pointlike source producing the bulk of the emission and a fainter, extended ($r > 1 $ kpc) emission. 

\subsection{Photoionization and shock models}
\label{sec:models}
We now complement the classification done in P17b with
 a more detailed analysis: we simultaneously consider the lines  
($H\alpha$, H$\beta$, [OI]$\lambda$6300, [OIII]$\lambda$5007, [NII]$\lambda$6583, [SII]$\lambda\lambda$6717,6731) commonly used to identify AGN  \citep{1987ApJS...63..295V,2006MNRAS.372..961K} and compare them to the predictions from  photoionization and shock models. 
In those cases where two components were required for the fit,  we considered the summed fluxes for the comparison with models.

We proceed as follows: 1. we selected only spaxels classified in P17b as either AGN, composite or LINERs; 2. for each spaxel we run   \texttt{NebulaBayes} \citep{2018ApJ...856...89T}, a python code that adopts a Bayesian approach to select the model optimally fitting the target emission line fluxes.

\begin{table}
	\begin{tabular}{ccccc}
		\hline\hline
		id & $\log P/k$ & $\log E_{\rm peak}$ & $12 + \log O/H$ & $\log U$\\
		\hline     
		JO135 & 6.6 &-1.2 & 9.16 & -2.49\\
		JO201 & 7.0 & -1.5  & 8.99 & -2.77\\
		JO204 & 7.0 & -1.5 & 8.99 & -3.06 \\
		JO206 & 6.6 & -1.5 & 8.87 & -3.06\\
		JW100 & 7.0 & -1.5 & 9.15 & -3.06 \\
		JO194 & 6.2 & -1.7 & 9.30 & -3.34\\
		\hline    
	\end{tabular} 
	\caption{Best-fit parameters for the nuclear AGN photoionization models.\label{tab:photmodels} }
\end{table}

\texttt{NebulaBayes} includes grids where constant gas pressure photoionization models are computed with the \texttt{MAPPINGS V} code, for HII regions and AGN. A full discussion of the assumptions and parameters of these models is given in  \citet{2018ApJ...856...89T}, we summarize here the  main aspects.

For HII regions, the ionizing continuum is defined by the \texttt{SLUG2} \citep{2015MNRAS.452.1447K} stellar population synthesis code, with  five metallicities ($Z$ = 0.0004, 0.004, 0.008, 0.02, 0.05).
For AGN, the ionizing continuum is described in \citet{2016ApJ...833..266T}, and is parametrized by the energy of the peak of the accretion disk emission ($E_{\rm peak}$),  the photon index of the
inverse Compton scattered power-law tail ($\Gamma$), and the proportion
of the total flux in the non-thermal tail ($p_{\rm NT}$). In the grid models, the latter two parameters are fixed ($\Gamma$=2, $p_{\rm NT}$=0.15).

For both HII regions and AGN, the other model free parameters are: the metallicity (12 + log O/H), the ionization parameter ($U$) and  the gas pressure ($\log P/k$), with $P/k \sim 2.4 n_e T$, see e.g. \citet{2018A&A...618A...6K}.

Considering the environment of these galaxies and the presence of outflows in the nuclear regions, it is important to understand what may be the contribution from shocks, and if shocks alone can produce the observed line ratios. 
As extensively discussed by \citet{2008ApJS..178...20A},  in the so-called fast shock models the cooling of the hot gas behind the shock front produces high energy photons which ionize the pre--shocked gas (precursor). When the shock velocity is $>$ 170 km s$^{-1}$, the contribution from the photoionized gas in the precursor starts to become increasingly important and both high and low ionization lines are present in the observed spectrum. Varying  the input model parameters, that is the pre--shock density, $n$, the shock velocity, $v_s$, the pre-shock transverse magnetic field, $B$, and the gas atomic abundances, it is possible to produce a wide range of emission line ratios, from HII-like regions to Liners and AGN. 

Since shock model libraries are not directly available in \texttt{NebulaBayes},
we adapted the  \citet{2008ApJS..178...20A} fast shock grids so that they could be used in \texttt{NebulaBayes}. From these grids, we selected the models with\footnote{We remind that the pre--shock density is not directly related to the electron density measured e.g. by the [SII]$\lambda\lambda$6716,6731 lines, giving the post--shock density \citep[see e.g.][]{1995ApJ...455..468D}}  $n$=0.1, 1, 10 cm$^{-3}$, for which solar ($n$=0.1, 1, 10 cm$^{-3}$) and 2x solar ($n$=1 cm$^{-3}$) abundances are available. 

 For each model type (shock, HII and AGN), \texttt{NebulaBayes} was run spaxel by spaxel, providing as output the best-fit model line ratios and the $\chi^2$: the optimal model was selected as the one giving the lowest $\chi^2$. 
We stress that different reasons may contribute to produce the wrong classification for a given spaxel, as for instance the uncertainties on the line measurements, the limited number of parameters in the models, and the fact that in many cases we may have at the same time a contribution from different ionizing mechanisms.

\subsection{Results: Shock vs. AGN}
\label{sec:shock2agn}

Fig.~\ref{fig:models} presents AGN and shock model grids overlaid on the observed spaxel line ratios; the latter are color coded with the distance from the galaxy center. For AGN models, we display those with $\log E_{\rm peak} =[-0,5,0,+0.5]$ around the best-fit value of the central spaxel, and for different values of $\log U$. The abundances in the nucleus derived from the NB fits are super-solar ($12+\log O/H > 9$) in the  AGN-dominated nuclei, and are Solar  ($12+\log O/H = 8.76$) outside.
For fast shock models, we plot a grid of varying velocities and magnetic field values, fixing the best fit density and metallicity.  

The maps of the best fit classification from \texttt{NebulaBayes}, compared with the classification derived as in P17b, are displayed in Fig.~\ref{fig:mapsKV2NB}.

Fast shock models in individual diagnostic diagrams can produce a wide range of line ratios, covering both the HII and AGN regions of the diagrams.
For the most extreme cases like JO135 and JO201 that have log [OIII]/H$\beta$ $\sim$ 1, in order to reach the observed line ratios the shock models require $v_{sh}>500 \, \rm km \, s^{-1}$
while in our case  the width of the strongest component is $\sigma_v \sim$ 100 km s$^{-1}$.
When $\log$ [OIII]$\lambda$5007/H$\beta$  is low ($<$ 0.5) the shock velocities do not need to be so extreme.
However, as shown in Fig.~\ref{fig:models}, for JO135, JO201, JO204, JO206, JW100, 
shock models produce either too high [OI]$\lambda$6300/H$\alpha$ ratios or too low [NII]$\lambda6583$/H$\alpha$ ratios, while AGN photoionization models reproduce all diagnostic ratios, as found for typical AGN by \citet{2008ApJS..178...20A}.

For JO194  the emission line ratios fall in the LINER side of the diagrams, with $\log$ [OIII]$\lambda$5007/H$\beta \le 0.2$,  where optical lines alone are not able to clearly separate between AGN and other ionization mechanisms \citep[see e.g.][]{2016MNRAS.461.3111B}. Around the central spaxel there is a very small region, whose size is few spaxels, where AGN models produce a lower $\chi^2$ than shock models. However, as displayed in Fig.~\ref{fig:mapsKV2NB} the line ratios move to the SF region of the diagrams already within 2 kpc; line ratios similar to what observed in the nucleus are also present up to $\sim$ 5 kpc. Considering the uncertainties due e.g. to the fact that here we do not consider mixed models where both SF and AGN or shocks contribute to the line ratios, we conclude that we can't confirm or discard the presence of the AGN; in P17b the AGN option was favoured, considering the high Chandra X--ray luminosity ($L_{\rm 0.3-8 keV} = 1.4 \times 10^{41}$ erg s$^{-1}$). Since we detect a strong extinction ($A_v > 2.5$) in the nuclear regions, it is possible that AGN emission is obscured by dust in the optical.

For JO175, line ratios are consistent with star formation, as well as with some of the shock models as suggested by the high [OI]$\lambda$6300/H$\alpha$ ratio, but not with an AGN. 

As a further test, we compare the observed, dust corrected L(H$\beta$) luminosity within $r_{\rm NLR}$  with the value derived from  the  \citet{2008ApJS..178...20A}  library, selecting models with $n=1$ cm$^{-3}$ and best-fit values in the central spaxel for $B$ and $v_{\rm sh}$. 
We estimate the maximum contribution from shocks to be negligible (< 3\%) for JO135,  < 20\% for JO194, JO201, JO204 and JO206,  < 40\% for JW100.

We conclude that, in agreement with P17b, the \texttt{NebulaBayes} results confirm that the central spaxels of all galaxies except JO175 are best fitted by AGN models, whose parameters are given in  Table~\ref{tab:photmodels}. For JO175, nuclear line ratios can be fitted either by SF or by shocks: HII-like [NII]$\lambda$6583/H$\alpha$ and [SII]$\lambda\lambda$6716+6731/H$\alpha$, but high [OI]$\lambda$6300/H$\alpha$, agree well with shock models (either fast or slow).

Finally, four galaxies (JO201, JO204, JO206 and JW100) also show AGN  line ratios in the circumnuclear regions ($< 5$kpc), with an expected decrease of the ionization parameter, and a decrease  of  $E_{\rm peak}$ consistent with an increasing contribution from HII ({\em composite}) regions. 
In fact, as discussed in \citet{2018ApJ...856...89T}, variations in $E_{\rm peak}$ may be due either to screening by gas and dust (hardening the ionizing continuum and thus increasing $E_{\rm peak}$), or to contamination from shock or HII regions (softening the continuum and thus decreasing $E_{\rm peak}$).

%=============== OIII maps

\begin{figure*}
\begin{tabular}{cc}
\includegraphics[width=.5\linewidth]{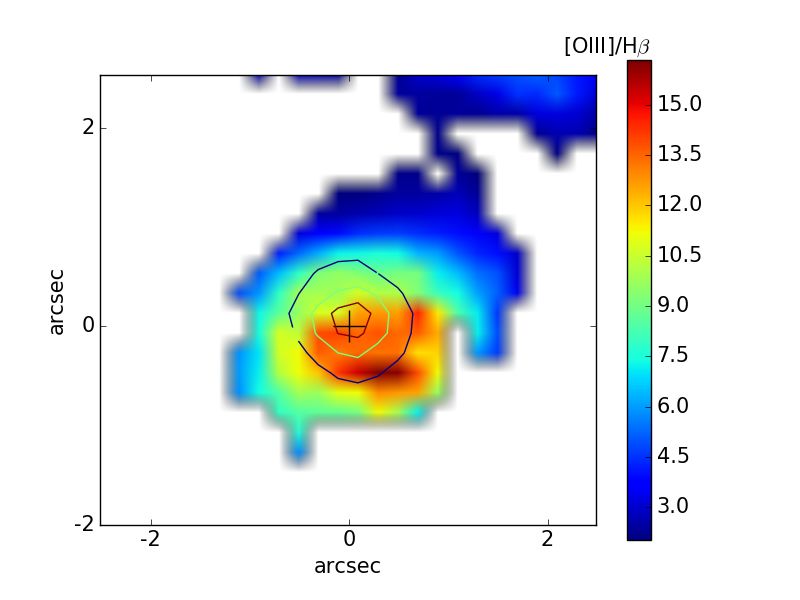} &
 \includegraphics[width=.5\linewidth]{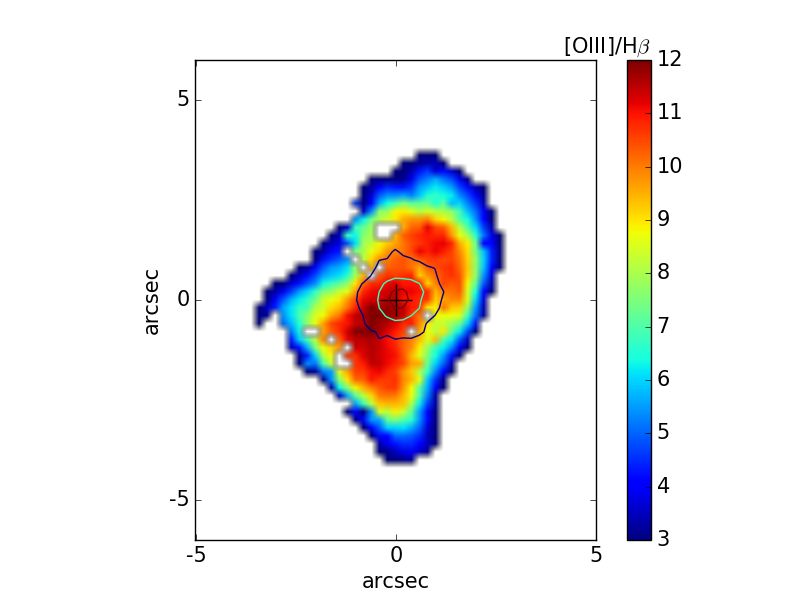}\\
\end{tabular}
	\caption{The contour maps showing emission at SN of $\sim$ 5 - 20 in the [Fe VII] $\lambda$6087 line are overplotted on the  
	[OIII]$\lambda$5007/H$\beta$ map for the nuclear AGN regions of JO135 ({\em left}) and JO201 ({\em right}).  The cross displays the position of the peak in [OIII] $\lambda$5007.
	\label{fig:maps_Fe7}}
\end{figure*}

\begin{figure*}
	\begin{tabular}{cc}
		\includegraphics[width=.5\linewidth]{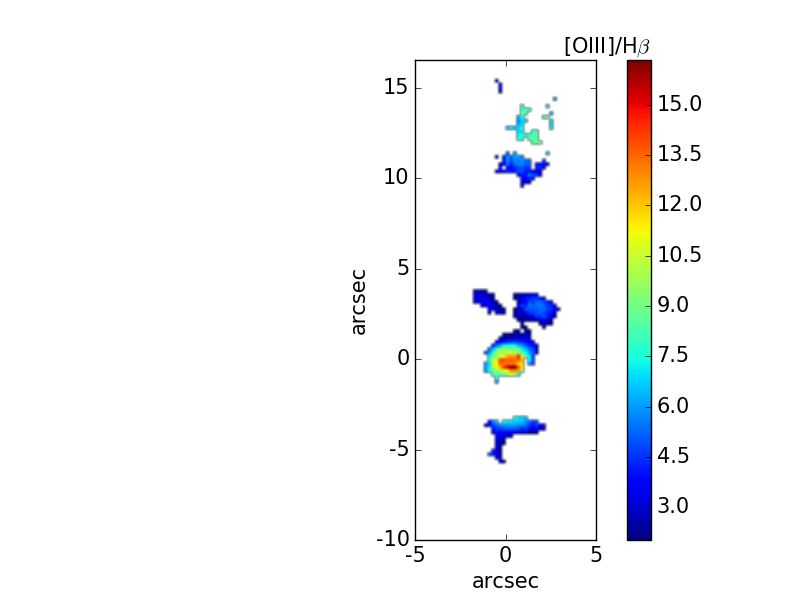}&
 \includegraphics[width=.5\linewidth]{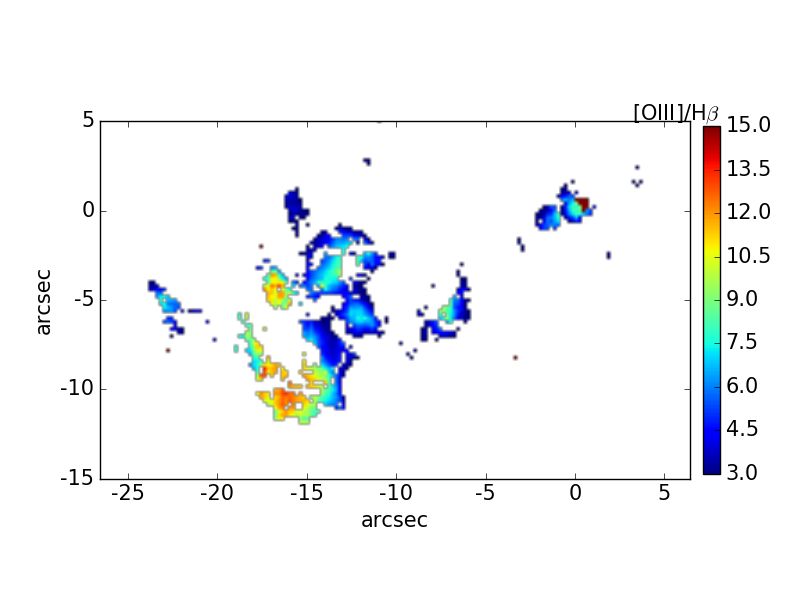}
	\end{tabular} 
	\caption{[OIII]$\lambda$5007/H$\beta$ maps showing the AGN-like extranuclear emission in JO135 ({\em left}) and JO204 ({\em right}). The coordinates are centered on the position of the central spaxel. \label{fig:maps_eelr}}
\end{figure*}

%===============================

\subsection{Coronal lines}
\label{sect:coronal}

The high ionization (coronal) line [Fe VII]$\lambda$6087 (Fig.~\ref{fig:spectra}, Fig.~\ref{fig:maps_Fe7}) is detected in the inner kpc region of JO201 and JO135, with a peak SN of $\sim$ 20, corresponding to [Fe VII]$\lambda$6087/H$\alpha$ $\sim$ 0.05. The weaker lines [Fe VIII]$\lambda$5721  and  [FeX]$\lambda$6375 (blended with [OI]$\lambda$6363) are also present in the same regions. 
We report the presence in JO206 of a faint feature in the [Fe VII]$\lambda$6087 spectral region, but the SN $<$ 3 is too low for a sure identification. 
 Both in JO135 and JO201, the [Fe VII]$\lambda$6087  line is characterized by a profile similar to that of low ionization lines, but it has a redshifted peak with respect to the Balmer recombination lines, in particular in JO201, suggesting that these lines are produced in different regions. This is confirmed by the fact that, compared to the other lines, its emission is concentrated (Fig.~\ref{fig:maps_Fe7}) in a region whose size is close to the size of the PSF (FWHM $\sim$ 1 arcsec). Adopting the definition of the NLR size presented before, we obtain  $r_{\rm NLR}$(FeVII) < 0.5 kpc, and $r_{\rm AGN}$(FeVII) $\sim$ 0.5 kpc (JO135), 1.2 kpc (JO201): it is therefore unresolved in JO135, while in JO201 there is a faint extended emission that could be however an artifact due to the PSF wings.  The intensities of coronal lines are not available in the \texttt{NebulaBayes} grids, since \texttt{MAPPINGS} does not accurately model these lines \citep{2016ApJ...824...50D}. Different models were  presented to reproduce coronal lines in AGN. 
 \citet{2019A&A...622A.146M} reported the presence of Fe coronal lines in a sample of AGN with outflows, observed with MUSE: they attributed them to the inner, optically thin and highly ionized regions of the outflows.
 \citet{1989ApJ...343..678K} attributed them to a low-density ($n_e$ $\sim$ 1 cm$^{-3}$ ) ISM heated by the AGN radiation, in a region whose size is similar or larger than the NLR ($\sim$ 1-2 kpc).
\cite{1997ApJ...487..122F,1997A&A...323...31K} showed that multi-component photoionization models, where emission lines are produced in an ensemble of clouds with different gas densities and distances from the center, are able to successfully fit the observed values: in particular, in these models  coronal lines are mostly produced by hot ($T\sim10^5$ K) gas heated by the AGN in the high-ionization inner regions \citep[see also][]{2010MNRAS.405.1315M}.  \cite{2017ApJS..232...11T} reported the existence of a correlation between [Fe VII]$\lambda$6087/H$\alpha$ and [OIII]$\lambda$5007/H$\beta$ in Seyfert galaxies in their sample, that they interpreted as an effect of a radiation pressure dominated environment, where Compton heating in the central regions triggers the production of coronal lines  \citep{2016ApJ...824...50D}. 
Consistently, both JO201 and JO135 show a high [OIII]$\lambda$5007/H$\beta$ (> 10) ratio in the inner regions where the emission of [FeVII] is observed. 
The absence of this line in the other galaxies may be due either to  orientation effects preventing us to see the inner regions, or to an intrinsic difference in the properties of the ionized gas.

\subsection{Extranuclear regions}
\label{sect:ExtNucleus}

AGN-like regions are detected in JO135 and JO204 \citep{2017ApJ...846...27G} up to $\sim$ 20 kpc from the nucleus (Fig.~\ref{fig:maps_eelr}),  with high values of [OIII]/H$\beta \sim 10$ (Fig.~\ref{fig:models}). This is typical of the so-called Extended Emission Line Regions (EELR)  \citep[see e.g.][]{2004AJ....127...90Y,2018MNRAS.480.5203M}, where the  gas  ionized by the AGN extends over scales of tens of kiloparsecs.  

In these regions the [SII]$\lambda$6716/$\lambda$6713 ratio, close to $\sim 1.4$, indicates low values of the electron density ($n_e < 50$ cm $^{-3}$), and the emission line widths are close to the instrumental value. The line ratios could be reproduced by shock models with $n \sim 0.1$, but the required shock velocity, $V_{\rm sh}$>400 km s$^{-1}$, is too high compared to the observed values: this rules out the presence of fast shocks \citep[see also][]{2009ApJ...690..953F}, and favors photoionization from the AGN (Fig.~\ref{fig:models}). 

In JO204, the required ionization parameter in the outer regions ($\sim 15$ kpc) is close to the value derived in the nuclear ($r < 2$ kpc)  regions ($\log U \sim -3$). A nearly constant ionization parameter  implies that the photon flux and the gas density should both decrease  as $r^{-2}$. This is consistent if there is a coupling between the radiation and the illuminated gas, as
in the case of radiation pressure mechanisms \citep{2018ApJ...856...89T}. The best-fit AGN models indicate for JO204 $\log P/k \sim$ 7 in the nucleus ($r \sim$ 1 kpc),   $\log P/k \sim$ 4.6 at 15 kpc: this would be consistent with $n_H \sim r^{-2}$, suggesting that the low-density gas in the ISM may be ionized by anisotropic radiation from the AGN.    
In order to verify if we can reproduce the observed H$\beta$ luminosities, we proceed as follows. For a gaseous cloud, the rate (photons s$^{-1}$) of ionizing photons required to produce  the observed, dereddened H$\beta$ luminosity (erg s$^{-1}$) is \citep{2006agna.book.....O}:

\begin{equation}
Q(H^0) = \frac{L(H\beta)}{h \nu_{H\beta}} \frac{\alpha_B(H^0,T)}{\alpha^{\rm eff}_{H\beta}(H^0,T)} \sim 2.09 \times 10^{12} \ L(H\beta)
\end{equation}

where $L(H\beta)$ is the observed, dereddened H$\beta$ luminosity (erg/s); in the Case B approximation  $\alpha_B(H^0,T)= 2.59 \times 10^{13}$ cm$^3$ s$^{-1}$,  $\alpha^{\rm eff}_{H\beta}(H^0,T) = 3.03 \times 10^{14}$ cm$^3$ s$^{-1}$ ($T\sim 10^4$ K). 

We can thus derive the rate of ionizing photons that should be emitted by the nucleus:
\begin{equation}
Q(H^0)_{\rm nuc}  = Q(H^0) \ \left(\frac{\Omega}{4\pi}\right)^{-1}
\end{equation}
where $\Omega$ is the solid angle covered by the extra-nuclear region, that is $\Omega = A/d^2$, for a region of area $A$ and  projected distance $d$ from the nucleus. From the $H\beta$ fluxes measured in the EENLR of JO204 at 15 kpc, we obtain\footnote{We neglect the factor due to the unknown projection, but here we are interested in order of magnitudes.} $Q(H^0)_{\rm nuc}$ $\sim$ $10^{54}$ ph/s: using this value to compute the ionization parameter, we obtain $\log U \sim -3$ for $n_H \sim 1$ $cm^{-3}$, in agreement with the value expected from photoionization models. Similar results are obtained for JO135.

%====== Figures: outflow section

\begin{figure*}
	\begin{tabular}{c}
		JO135 \\
		\includegraphics[width=.9\linewidth]{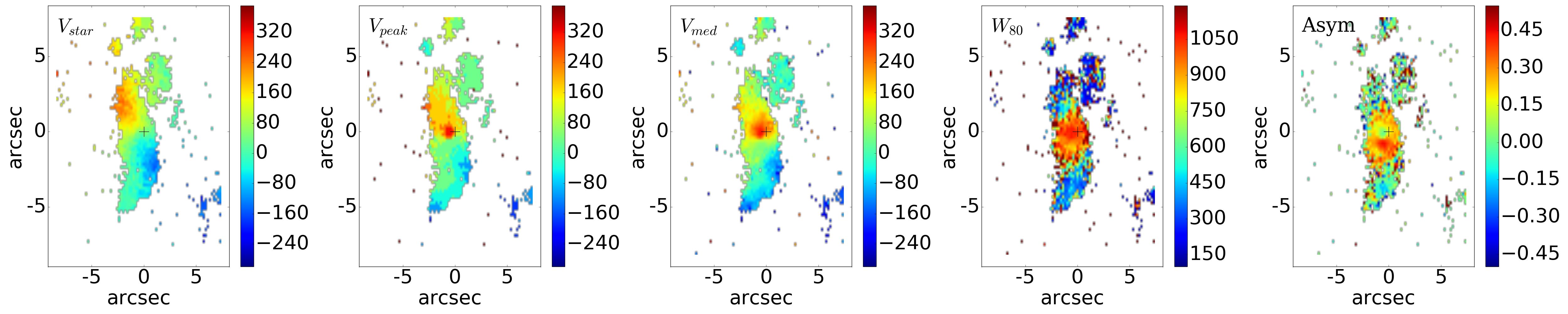}\\
		JO201 \\
		\includegraphics[width=.9\linewidth]{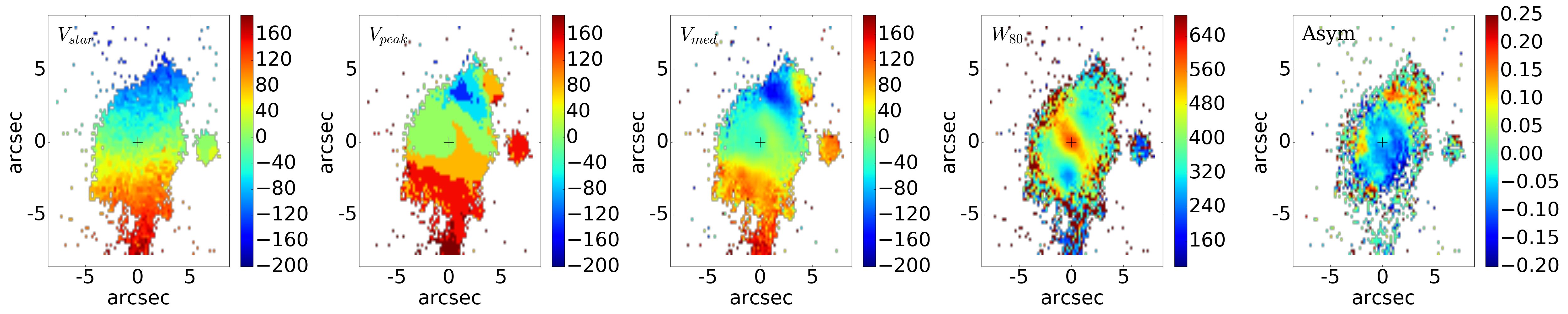} \\
		JO204\\
		\includegraphics[width=.9\linewidth]{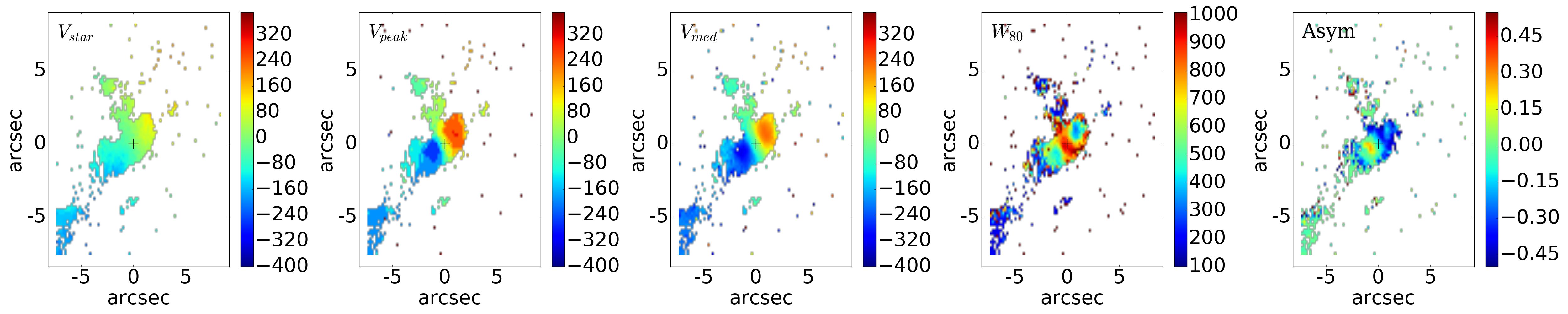}\\
		JW100\\
		\includegraphics[width=.9\linewidth]{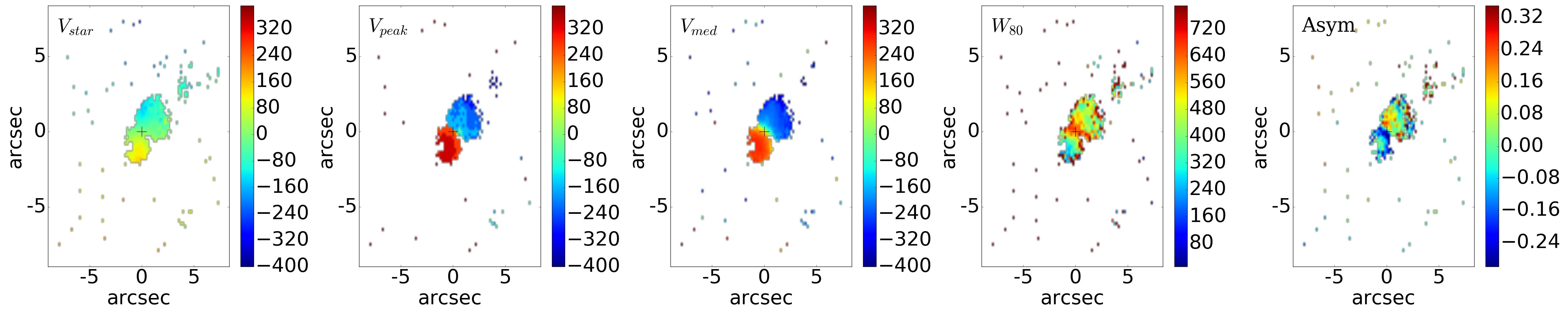}\\
	\end{tabular} 
	\caption{For each galaxy hosting an AGN, the plots display the spatial distribution  of the stellar velocity [km s$^{-1}$] ({\em first panel}) and of the following parameters (see text for details) derived from the [OIII] line: peak and median velocity [km s$^{-1}$];
	$W_{\rm 80}$ [km s$^{-1}$]; asymmetry ({\em right panel}). In each plot, North is up and East is at left.\label{fig:npars}}
\end{figure*}

\begin{figure*}
	\begin{tabular}{c}
		JO135 \\
		\includegraphics[width=.9\linewidth]{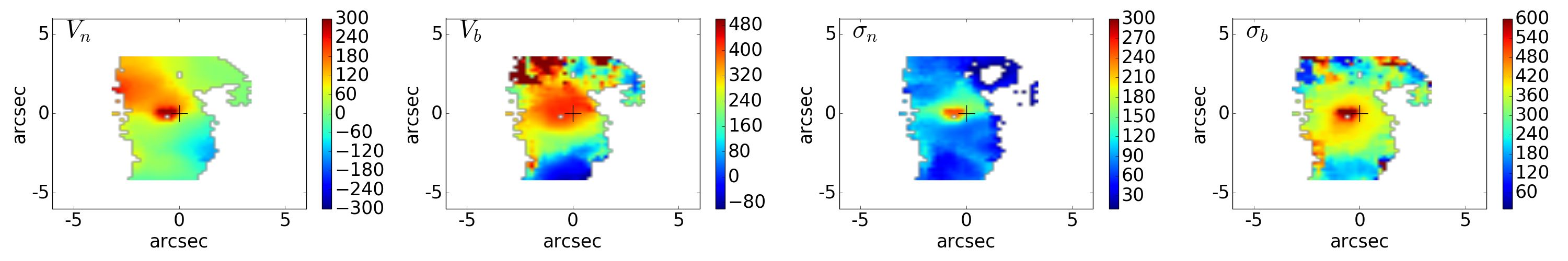}\\
		JO201 \\
		\includegraphics[width=.9\linewidth]{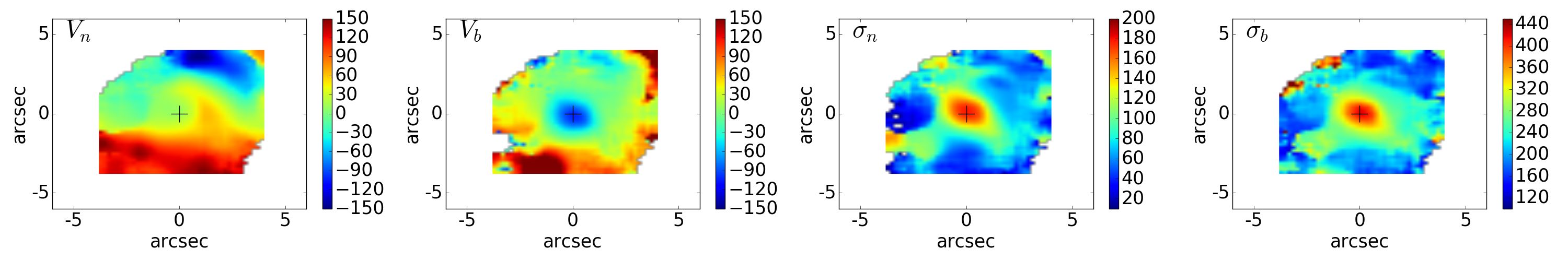} \\
		JO204\\
		\includegraphics[width=.9\linewidth]{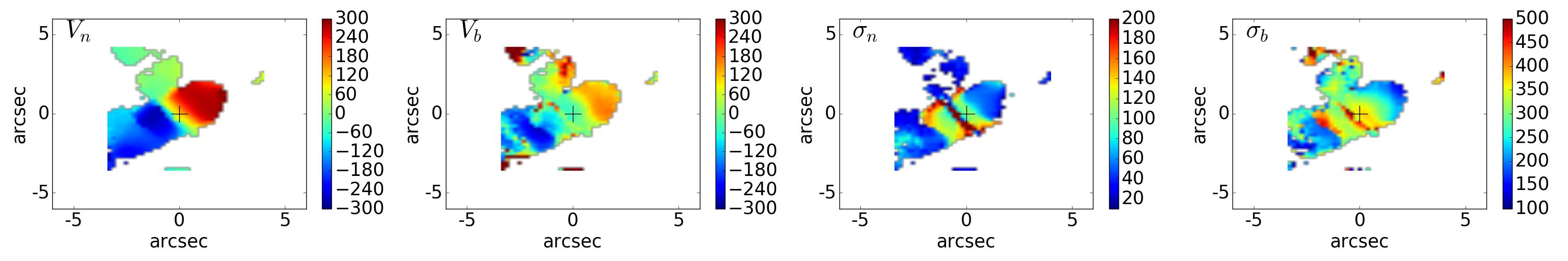}\\
		JW100\\
		\includegraphics[width=.9\linewidth]{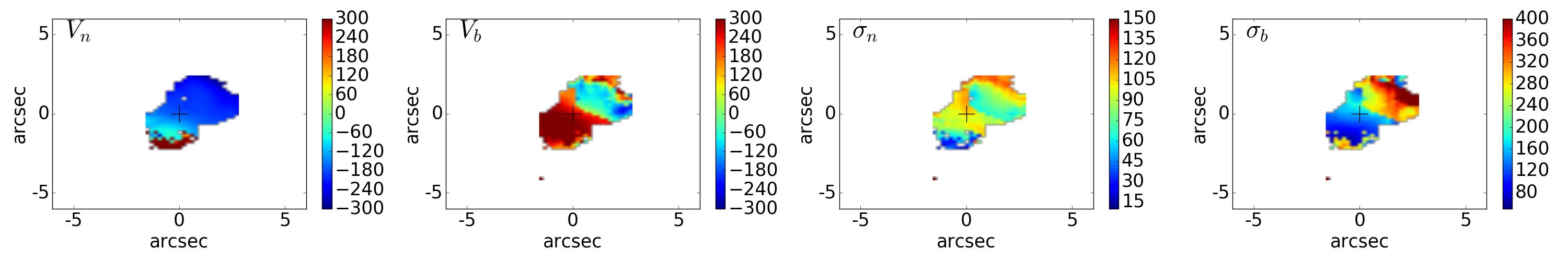}\\
	\end{tabular} 
	\caption{Spatial distribution of velocity and velocity dispersion from [OIII]$\lambda$5007 for galaxies with two components.\label{fig:fpars}}
\end{figure*}

%======================================================

\section {Gas properties: disk and outflow components}
\label{sect:Outflows}

As discussed before,
in four galaxies (JO201, JO204, JW100 and JO135) emission lines in the circumnuclear regions are characterized by complex profiles, that require at least two Gaussian components to be fitted. In order to disentangle the contribution to the emission lines from gas in disk and other components (e.g. outflows), we focus on the  [OIII] $\lambda$5007 line. Compared to the  H$\alpha$+[NII] lineset used in P17b, [OIII] is more suitable for this analysis as it is not affected by the presence of other nearby emission lines (the [NII] doublet in the case of $H\alpha$) or by possible residual broad components in the inner AGN regions, as it may be the case for permitted lines, and better traces the ionized gas in the outflows \citep[see e.g.][and refs.]{2014ApJ...795...30B}.
To this end, we selected a region of $\sim$ 10\arcmin x 10\arcmin around the center of each galaxy and fitted the [OIII]$\lambda$5007 line, using  the  functions available in the Python library \texttt{lmfit}.
For each spaxel, we made the fit adopting both one and two Gaussian components. The two component solution was chosen if it gave an appreciable improvement to the fit compared to the one component solution: based on visual inspection, we defined this condition as  $\chi^2_{n=1}$ > 1.5 $\chi^2_{n=2}$, $\chi^2$ being the chi-square of the fit \citep[see][for a similar approach]{2012MNRAS.426.1574D}, and in addition we requested that the flux in each component must be at least 10\% of the summed flux. We also discarded those fits where S/N([OIII]) $<$ 5, where the signal $S$ is the  total line flux and the noise $N$ is the standard deviation of the fitting residuals. 

From the velocities measured at 10\%, 50\% and 90\% of the cumulative flux percentiles, we used the definitions in \citet{2013MNRAS.436.2576L} to estimate the  parameters introduced by \citet{1985MNRAS.213....1W}:  the peak velocity of the [OIII] line ($v_{\rm pk}$), the median velocity ($v_{\rm 50}$),  the width $W_{\rm 80} = v_{\rm 90} - v_{\rm 10}$ and the asymmetry $A_{\rm sym} = \frac{(v_{\rm 90} - v_{\rm 50}) - (v_{\rm 50} - v_{\rm 10}) }{W_{\rm 80}}$. In this definition,  positive/negative values of $A_{\rm sym}$ indicate red/blue  asymmetric lines. 
We used the fit to model the line profile and compute these  parameters. In this way we do not assign any physical meaning to the decomposition, using the fit only to reduce the effect of the noise on the estimate of the profile parameters \citep{2013MNRAS.436.2576L,2014MNRAS.441.3306H, 2016AA...585A.148B}. 

The spatial distribution of these parameters for the galaxies showing two emission line components (JO135, JO201, JO204 and JW100) is displayed in Fig.~\ref{fig:npars}, where the velocity measured for stars is also displayed as reference. 

For a more quantitative analysis, we then analyzed (Fig.~\ref{fig:fpars}) the velocity and velocity dispersion of the two fitted components, bearing in mind that  the fit may be degenerate, in particular when the components are close, or in the outer regions where the line is fainter and noise may introduce spurious features.  

\begin{figure*}
	\begin{tabular}{cc}
		JO135 & JO201\\
		\includegraphics[width=.5\linewidth]{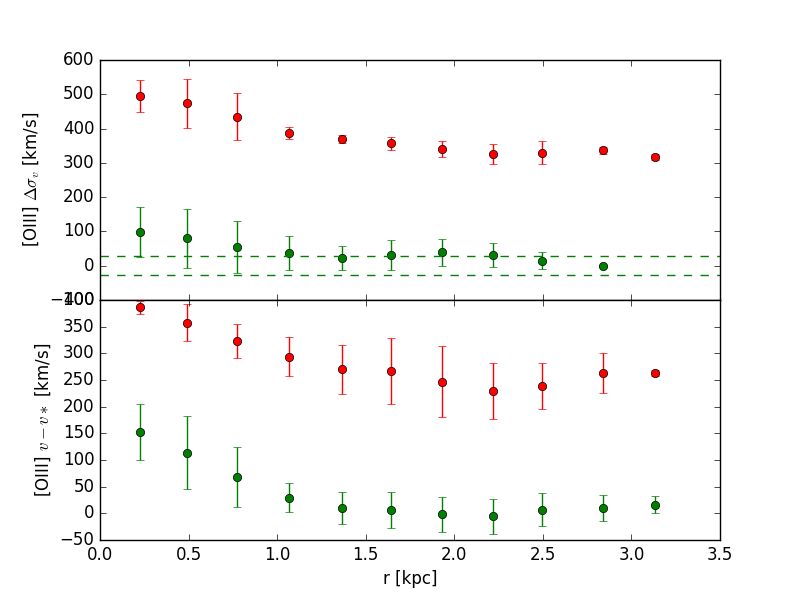}& 
		\includegraphics[width=.5\linewidth]{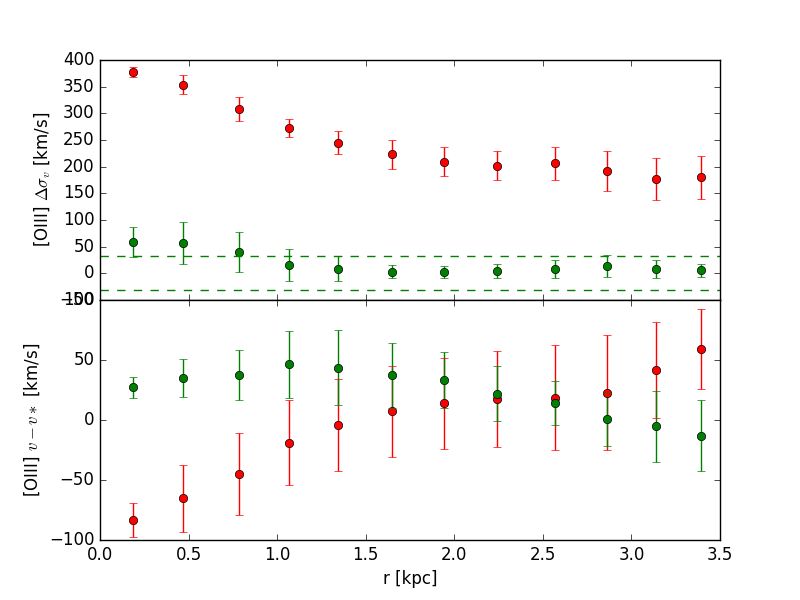}\\
		JO204 & JW100\\
		\includegraphics[width=.5\linewidth]{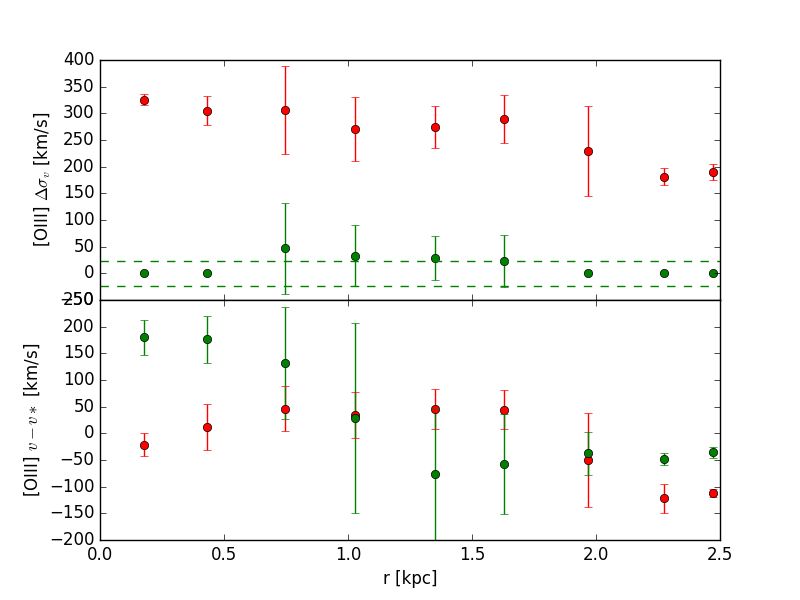} &
		\includegraphics[width=.5\linewidth]{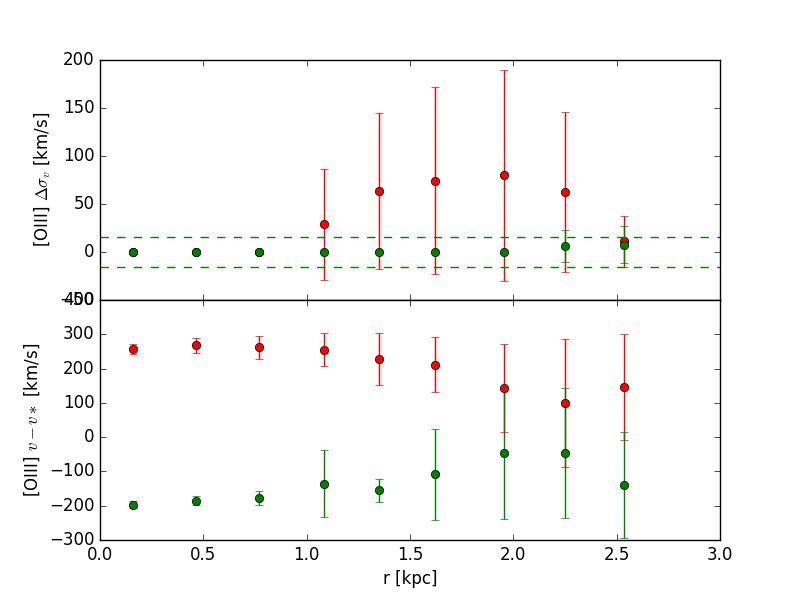}\\
	\end{tabular} 
	\caption{Velocity and velocity dispersion of the narrow (green) and broad (red)
		components of [OIII]$\lambda$5007. 
		The velocity of the stars is subtracted from the velocity, while $\Delta \sigma_v$ is the velocity dispersion from which the stellar velocity dispersion was subtracted in quadrature.
		For the velocity dispersion, a value of 0 was assigned if the velocity dispersion was lower than the stellar velocity dispersion. \label{fig:vvd}}
\end{figure*}

In the presence of an outflow we expect to see a clear separation in both velocity and velocity dispersion between a (primary) component, kinematically dominated by the gravitational potential traced by the stellar component, and a secondary (slightly broader) component with a velocity offset showing its non-gravitational origin \citep{2016ApJ...817..108W,2016ApJ...819..148K}. To check if this is the case, Fig.~\ref{fig:vvd} shows the radially binned values of the deviations from the stellar values of the velocity and velocity dispersion.

As discussed in  \citet{2010ApJ...708..419C}, the observed line profiles can be explained by a combination of biconical outflows and extinction from dust in the inner galaxy disk.  Biconical outflows are most often observed as broad, blueshifted components in the emission lines as the redshifted, receding part of the outflow more likely lies behind the galaxy disk and is thus suppressed by dust. 
In some cases, instead, red asymmetric lines are observed: this can still happen when the inclination of the disk is such to hide the approaching part of the outflow. 
In addition, \citet[][see their Fig.~15]{2015ApJ...806...84L}  proposed a model where dust is embedded in the outflowing clouds: in this case,
blueshifted clouds at small distances from the center preferentially show their non ionized face and are thus fainter compared to redshifted clouds, showing instead their ionized face.
At large projected distances, an increasing fraction of the ionized face is visible in the blueshifted clouds, which are then brighter.

Fig.~\ref{fig:npars}, Fig.~\ref{fig:fpars} and Fig.~\ref{fig:vvd} demonstrate that the four galaxies with double components have an outflow, and that their outflow properties are quite diverse, as we discuss in the following.

\begin{table*}
	\begin{tabular}{ccccccccccc}
		\hline\hline
		id & $r_{\rm 50}$ & $r_{\rm 90}$ & $\log E_{\rm kin}$ & $\bar{v}_{\rm out}$&$\log t_{\rm out}$  & $\log M_{\rm out}$  & $\log \dot{M}_{\rm out}$ & $\log \dot{E}_{\rm kin}$  & $\log L_{\rm AGN}$\\
		& kpc & kpc & erg & km s$^{-1}$ & yr & $M_\odot$ & $ M_\odot$ yr$^{-1}$ & erg s$^{-1}$ & erg s$^{-1}$ \\
		\hline     

JW100 & 1.1 & 2.1 & 50.5 & 248& 6.9& 4.40 & -2.53& 36.0 & 43.9\\
JO201 & 1.1 & 2.4 & 53.7 & 261& 6.9& 5.99 & -0.95& 39.2 & 45.4\\
JO204 & 0.7 & 1.2 & 52.2 & 317& 6.6& 4.27 & -2.31& 38.1 & 43.8\\
JO135 & 0.9 & 1.5 & 53.7 & 544& 6.4& 5.27 & -1.16& 39.7 & 44.8\\

		\hline    
	\end{tabular} 
	\caption{Outflow properties derived as described in the text.\label{tab:outflow}}
\end{table*}

\subsection{Outflows properties for individual galaxies}

\begin{description}
\item[JO135]: within a radius $\sim$ 1 kpc from the center, it presents a quite broad [OIII]$\lambda$5007 ($W_{80}$ $\sim$ 1000 km s$^{-1}$), which is asymmetric in the red (Fig.~\ref{fig:npars}): in the same region, there is an increased dust extinction ($A_v \sim 2$), that also extends in the outer regions  (Fig.\ref{fig:dpars}). The presence of dust could point to the model proposed by \citet{2015ApJ...806...84L}, to explain the  dominant redshifted component in the outflow. 
 Outside this region, the velocity pattern follows the stellar one. This can also be seen in the radial plots of the fitted components (Fig.~\ref{fig:vvd}), where at $r>$ 1 kpc the velocity and velocity dispersion of the narrow component are very close to the stellar values. 

\item[JO201]: a broader component ($W_{80}$ $\sim$ 600 km s$^{-1}$) is detected in the inner kpc, with a small blue asymmetry. In the same region,  we measure an increase of the dust extinction ($A_v \sim 1$), as well as of the electron density ($n_e \sim 10^3$ cm$^{-3}$).
The velocity and velocity dispersion of the narrow component agree with the stellar values: we therefore identify the narrow component with gas in the disk and the broad blueshifted component with the outflow.
The low asymmetry in [OIII] may imply a spherical or wide-angle outflow  \citep{2013MNRAS.436.2576L}, allowing to see the inner outflow regions and hence the [FeVII] $\lambda$6087 line, as discussed in Sect.\ref{sect:coronal}. 

\item[JO204]: both the peak and median velocities differ from the stellar velocities \citep[see also][]{2017ApJ...846...27G} within a region of size $\sim$ 2 kpc. In a narrow central strip
around the center, emission line profiles are complex (see Fig.\ref{fig:spectra}), thus producing higher values of $W_{80}$ ($\sim$ 900 km s$^{-1}$). Outside this strip, we identify two regions, one (NW) with blue asymmetric lines and one (SE) with red asymmetric lines. From the two-component fits, we find a narrow component, where $\sigma_v$ is close to the stellar values, and a significantly broader, but fainter, one ($\sigma_v \le 500$ km s$^{-1}$). Since up to a distance of 1 kpc both components have a velocity that is significantly different than the stellar value, we interpret both of them as produced by the two sides of a biconical outflow; at larger distances ($>$ 1 kpc) the gas is most likely dominated by the disk component.

\item[JW100]:  we detect two line components,  both of which are  narrow ($\sigma_v < 200$ km s$^{-1}$), blue and redshifted, with a velocity offset compared to the stellar velocity of $v \sim \pm 200$ km s$^{-1}$ up to a radius $r$ $\sim$ 2.5 kpc. 
The two components are emitted from  distinct regions, with the exception of a small area around the center  where a double-peaked profile is observed. As for JO204, we interpret this as a biconical outflow, extending to a distance of $\sim$ 2 kpc.

\end{description}

\subsection{Outflow: size, mass and energy}

From the {\em secondary} (broader) components, we computed  the radius containing a fraction $f$ (f = 50\% and 90\%) of the total broad [OIII] flux:  $\sum_{r<r_{\rm f}} F_{\rm [OIII], broad}$ = f $\sum_{r<r_{\rm max}} F_{\rm [OIII], broad}$, $r_{\rm max}$ being the outer radius where the presence of the second component is significant. We adopted $r_{90}$ as the outflow size as it agrees with a dynamic definition of the outflow size \citep{2016ApJ...819..148K}, that is the radius where  the velocity and velocity dispersions of the outflow component start to decline and  approach  the stellar values. 

The mass related to the outflow is \citep{2015A&A...580A.102C}
$M_{\rm out} = 8 \times 10^7 M_\odot C \frac{L[OIII]}{10^{44} erg s^{-1}} \left(\frac{<n_e>}{500 cm^{-3}}\right)^{-1}$, with $C=\left<n_e\right>^2/\left<n_e^2\right> \sim ~ 1$. A typical density $\left<n_e\right> = 500$ cm $^{-3}$ was assumed.
The outflow kinetic energy is $E_{\rm kin} = \frac{1}{2} M_{\rm out} v^2_{\rm out}$, $v_{\rm out}$ being the bulk velocity of the outflow. As discussed e.g. in \cite{2016ApJ...833..171K}, different choices to estimate $v_{\rm out}$ were made in the literature, reflecting different strategies to take into account geometrical effects (e.g. projection, opening angle of the outflow). Here we adopt the approximation \citep{2016ApJ...833..171K, 2018NatAs...2..198H} $v^2_{\rm out} = v^2_{\rm rad} +  \sigma^2$, where $v_{\rm rad}$ is the measured radial velocity and $\sigma$ the [OIII] velocity dispersion, corrected  \citep{2016ApJ...833..171K} for the contribution from the gravitational potential by subtracting in quadrature the stellar velocity dispersion. The mass outflow was derived summing on all spaxels within $r_{90}$ the contributes from both line components since, with the possible exception of JO201, we were not able to unambiguously separate the disk and outflow contributions to [OIII]. 

From the mean bulk velocity and the size of the outflow we can then compute the outflow lifetime, $t_{\rm out} = r_{\rm out}/\bar{v}_{\rm out}$, the  outflow mass rate, $\dot{M}_{\rm out}=M_{\rm out}/t_{\rm out}$, the energetic rate, $\dot{E}_{\rm kin}=E_{\rm kin}/t_{\rm out}$ and the 
outflow efficiency, $\eta =  \dot{E}_{\rm kin}/L_{\rm AGN}$ where $L_{\rm AGN}$ is the bolometric luminosity. 
In \citet{2016ApJ...833..171K} the bolometric  AGN luminosity was computed as $L_{\rm AGN} = 3500 L_{\rm[OIII]}$ erg/s \citep{2004ApJ...613..109H},   $L_{\rm[OIII]}$ being the dust uncorrected luminosity, for comparison with other literature samples. We adopted the same choice and  used the [OIII] luminosities in Table~\ref{tab:props}, that include both emission line components.

The results obtained from the above analysis are displayed in Table~\ref{tab:outflow}.
We emphasize all the uncertainties  related both to the outflow kinetic energy and the estimate of the bolometric luminosity. Nevertheless, the outflow mass rates and kinetic energies that we obtain are comparable with the values obtained by \citet{2016ApJ...833..171K} in a sample of AGN having similar [OIII] luminosities
($L_{\rm[OIII]} < 10^{42}$ erg s$^{-1}$). Consistently with these results, we derive low efficiencies, $\eta \ll 0.01\%$,  suggesting that the outflow is not able to impact the host galaxy environment on large scales. As discussed by e.g. \citet{2019MNRAS.482..194B}, moderate luminosity AGN may however suppress the star formation in the inner few kpc. For the galaxies in our sample, the outflow lifetime derived above is $t_{\rm{out}} \sim 10^7$ yr and the outflow velocity is $v_{\rm{out}} \sim 300$ km s$^{-1}$: this corresponds to a distance of  $\sim$ 3 kpc up to which the outflow can propagate from the center, in agreement with the outflow size ($r_{90}$ in Table~\ref{tab:outflow}). Evidence for star formation suppression around the AGN in JO201, based on NUV and CO data, will be presented in George et al. (2019, submitted).
For comparison, from literature the mass outflow  and energy rates in the brightest AGN ($L_{\rm[OIII]} > 10^{43}$ erg s$^{-1}$) can be as high as $10^4$ $M_\odot$ yr$^{-1}$ and $\log \dot{E}_{\rm kin} \sim 10^{45}$ erg s$^{-1}$ \citep{2013MNRAS.436.2576L}, thus being able to impact on much larger scales in the host galaxy environment.

\section{Conclusions}
\label{sect:conclusions}

In this paper we have carried out a detailed investigation of the seven  jellyfish galaxies presented in P17b, where based on the [NII]/$\rm H\alpha$ ratio it was found that at least five of them (JO201, JO204, JO206, JO135 and JW100) host an AGN. We first performed a detailed comparison with photoionization and shock model taking into account several diagnostic diagrams simultaneously. We concluded that while shock models can play a role in the ionization of the gas, AGN models are required to explain the line ratios observed in the nuclear regions of these five galaxies: this conclusion is corroborated by an analysis of the $\rm H\beta$ luminosity. The presence of iron coronal lines in the nuclei of JO201 and JO135 indicates the existence of hot ($T \sim 10^5$K) gas heated by the AGN. 
JO204 and JO135 also present Extended Emission Line Regions of $>10$ kpc that are ionized by the AGN. In JO194, that was classified as a LINER in P17b, line ratios in the central spaxels are better reproduced by an AGN model, though shock models may also marginally reproduce the observed ratios.
Finally, in JO175 the [NII]$\lambda$6583/H$\alpha$ ratio is typical of star forming regions, but we still observe a high [OI]$\lambda$6300/H$\alpha$ ratio that could point to the presence of shocks.

We then focused on the [OIII]$\lambda$5007 line profile, which is a good tracer of the presence of outflowing gas. Four of the galaxies hosting an AGN present complex emission line profiles, with at least two Gaussian components with different line widths that testify the presence of AGN outflows. We have studied the properties and energetics of the outflows, with results comparable to what reported in other AGN  of similar luminosity ($L_{\rm[OIII]} < 10^{42}$ erg s$^{-1}$), with outflows. Finally, we derived conclusions on possible AGN feedback effects on the circumnuclear regions ($\sim 3$ kpc).  

This study confirms on much more solid ground the conclusions from P17b regarding the presence of an AGN in several GASP jellyfish galaxies with the longest tails, that suggests a causal connection between ram pressure stripping and AGN activity. Moreover, in this work we demonstrate the presence of outflows and derive their properties. The current sample is too small to draw final conclusions regarding the AGN-ram pressure connection, and a detailed analysis of the whole GASP sample will be presented in a forthcoming paper, investigating the evidence for AGN activity in the other jellyfish galaxies with long tails but also for different stripping stages and as a function of the galaxy mass.

\section*{Acknowledgements}
This work is based on observations collected at the
European Organisation for Astronomical Research in the Southern
Hemisphere under ESO program 196.B-0578. 
We acknowledge financial support from PRIN-SKA ESKAPE-HI (PI L.Hunt). Y.J. acknowledges financial support from CONICYT PAI (Concurso Nacional de Inserci\'on en la Academia 2017) No. 79170132 and FONDECYT Iniciaci\'on 2018 No. 11180558.
We acknowledge the usage of the following Python libraries: \texttt{AstroPy}, \texttt{lmfit}, \texttt{NebulaBayes}, \texttt{lineid\_plot}, \texttt{mpdaf}.
%%%%%%%%%%%%%%%%%%%%%%%%%%%%%%%%%%%%%%%%%%%%%%%%%%



\bibliographystyle{mnras}
\bibliography{paper} % if your bibtex file is called example.bib

%%%%%%%%%%%%%%%%%%%%%%%%%%%%%%%%%%%%%%%%%%%%%%%%%%

%%%%%%%%%%%%%%%%% APPENDICES %%%%%%%%%%%%%%%%%%%%%

% Don't change these lines
\bsp	% typesetting comment
\label{lastpage}
\end{document}